\documentclass[12pt]{article}
\usepackage{algorithm}
\usepackage{algpseudocode}
\usepackage{amsmath}
\usepackage{amssymb}
\usepackage{amsthm}
\usepackage{authblk}
\usepackage{bm}
\usepackage{comment}
\usepackage{caption}
\usepackage[top=0.5 in,bottom=0.5 in, left=0.5 in, right=0.5 in]{geometry}
\usepackage{graphicx} 
\usepackage{multirow}
\usepackage[round]{natbib}   
\usepackage{subfig}
\usepackage{xcolor}
\usepackage{setspace}
\makeatletter
\renewcommand*\env@matrix[1][*\c@MaxMatrixCols c]{%
  \hskip -\arraycolsep
  \let\@ifnextchar\new@ifnextchar
  \array{#1}}
\makeatother

\algrenewcommand\algorithmicrequire{\textbf{Input:}}
\algrenewcommand\algorithmicensure{\textbf{Output:}}

\newtheorem{theorem}{Theorem}
\newtheorem{proposition}{Proposition}

\usepackage{abstract}
\title{Collaborative Optimal Design of Controlled Experiments with Subject Covariates}

\author[1]{William Fisher}
\author[2]{Minshen Xu}
\author[3]{Qiong Zhang\thanks{Corresponding author: qiongz@clemson.edu}}
\author[2]{Lulu Kang}
\author[4]{Xinwei Deng}
\affil[1]{JMP Statistical Discovery LLC}
\affil[2]{Department of Mathematics and Statistics, University of Massachusetts Amherst}
\affil[3]{School of Mathematical and Statistical Sciences, Clemson University}
\affil[4]{Department of Statistics, Virginia Tech}
\vspace{-1.5em}
\date{}

\begin{document}
\vspace{-1in}
\maketitle
\vspace{-24pt}
\begin{abstract}
Modern experimental design problems often encounter multiple controlled experiments with subjects participating in more than one experiment. 
It is of interest to consider the optimal design of allocating subjects to treatment or control for these experiments especially when subject covariate information is available. 
In this work, we propose an optimal allocation method to enable accurate estimates of treatment effects for all experiments. 
Deriving the precision matrix of the treatment effects and using \textit{D-optimality} as our allocation criterion, we demonstrate the advantage of \textit{collaboratively} designing and analyzing multiple experiments over traditional \textit{independent} design and analysis, and propose efficient algorithms to design construction of the D-optimality problem for collaborative experiments. 
The performance of the proposed methods is examined by simulation studies and a case study involving online education experimentation.
\end{abstract}

\paragraph{Keywords:} Design Efficiency; Fast Algorithm; Mixed Effects Model; Online Experimentation; Semi-definite Programming;  

\section{Introduction} \label{intro}
Controlled experiments are commonly used to evaluate a proposed treatment by comparing outcomes between a treatment group and a control group. 
For example, a pharmaceutical company may assess a new drug by randomly assigning subjects to receive the drug or a control condition and comparing their blood pressure. 
More broadly, controlled experiments support analysis and decision-making in healthcare \citep{lyngbakken2020pragmatic}, transportation \citep{skippon2016experience}, and business \citep{koning2022experimentation}. 
In practice, research groups and organizations may conduct multiple controlled experiments with subjects participating in more than one experiment \citep{xu2015infrastructure,ostrow2016assessment}. 
Shared subject-level characteristics can induce dependence among responses across experiments, while observed subject covariates may also affect the outcomes. 
Designing experiments separately \emph{overlooks} this dependence, but accounting for it in the analysis alone does not ensure that its potential efficiency gains are realized. 
Allocations that balance covariates within each experiment may still be poorly coordinated across experiments, limiting the gains from joint analysis. 
The central design challenge is therefore to coordinate treatment allocations across experiments while maintaining covariate balance within each experiment, so that shared-subject information can improve the precision of treatment-effect estimates.

Optimal design of experiments \citep{pukelsheim2006optimal} provides a framework for constructing such designs to increase the precision of treatment-effect estimates when experiments are analyzed jointly. 
A commonly used criterion is \textit{D-efficiency} \citep{pukelsheim2006optimal} defined by
the determinant of the precision matrix of the estimated treatment effects for $K$ experiments. 
This quantity is inversely related to the volume of the treatment-effects confidence region; design with higher D-efficiency yield smaller confidence regions and therefore more precise estimates. 
Maximizing this criterion is the \textit{D-optimality} problem. 
By modeling experiments jointly and incorporating dependence across experiments, we can derive the precision matrix of the estimated treatment effects as a function of the allocations in the $K$ experiments. 
Solving the D-optimality problem then determines allocations that balance covariates within experiments  \citep{li2018balancing} while also promoting orthogonality of allocations across experiments to improve overall precision. 
However, the resulting optimization problem is challenging because the precision matrix involves the binary treatment allocations across experiments. 
Optimizing each experiment separately does not generally optimize the joint criterion, whereas directly searching over all allocations of $N
$ subjects and $K$ experiments requires considering a space of size $2^{NK}$. 
Practical design methods must therefore account for this complexity without incurring the computational cost of exhaustive joint optimization.

Related works have considered designing controlled experiments in the presence of covariates and dependence structures. 
\cite{bertsimas2015power} studied a single controlled experiment with covariates and proposed a parametric mixed-integer linear programming approach to balance covariates by minimizing discrepancies in the first and second moments between treatment and control. 
\cite{bhat2020near} considered offline and online allocation for a single experiment to maximize treatment-effect precision, using a
semi-definite program (SDP)
relaxation-based randomized algorithm for the offline problem and dynamic programming for the online problem. 
\cite{zhang2022min} studied two-armed trials with covariates for personalized decision-making in precision medicine; their setting included treatment-covariate interactions, and they optimized a surrogate objective. 
\cite{li2023optimal} extended \cite{zhang2022min} to multiple treatments composed of binary factors interacting with covariates and used a mixed-binary SDP to solve an E-optimality problem \citep{boyd2004convex}. \cite{zhang2022locally} studied \textit{network} A/B testing, where dependence arises through an underlying social network and proposed locally optimal designs that balance covariates and network connections. Rerandomization \citep{morgan2012rerandomization} and matching/weighting methods \citep{imai2014covariate} are also widely used to improve covariate balance in controlled experiments. 
The work most closely related to our setting is \cite{zhang2024collaborative}, which considered two experiments with shared subjects without covariates. 
\cite{zhang2024collaborative} proposed a collaborative analysis framework, and showed efficiency gains relative to analyzing experiments separately.
They identified collaborative design of multiple experiments with covariates as an open problem, which we address here. 

Our contributions connect the statistical benefits of collaborative experimentation with practical design construction. 
First, we derive the precision matrix of the treatment-effect estimators under a mixed-effects model, making explicit how subject covariates, shared subject effects, and treatment allocations jointly determine estimation precision. 
Second, we establish bounds on joint precision and individual treatment-effect variances. 
These results identify covariate balance and cross-experiment orthogonality as sufficient conditions for attaining the optimal bounds and characterize when collaborative analysis improves upon separate analysis. 
In particular, using the same allocation in every experiment eliminates the variance advantage of collaborative analysis, demonstrating why coordinating allocations matters.
Third, we develop two computational approaches for constructing efficient collaborative designs. 
A sequential greedy approach accounts explicitly for previously selected allocations, with each binary quadratic subproblem solved either by integer programming or by semi-definite relaxation and randomized rounding. 
A second approach solves a single semi-definite program and generates allocations for all experiments from its solution, reducing the computational burden. 
These approaches provide alternatives for settings with different computational and randomization requirements.

\section{Collaborative Design of Multiple Experiments} \label{sec: statmod}

To derive appropriate design optimality criterion, we model dependence between responses from different experiments recorded on the same subject using a subject-specific random effect. 
Beyond inducing correlation, this random effect also captures subject-level variation in the response that is not explained by observed covariates. 
We assume no treatment-covariate interactions within experiments and no treatment interactions across experiments.
We also assume that responses are measured on a common scale across experiments and, at the design stage, that every subject participates in every experiment. 
Under these assumptions, we jointly model the $K$ experiments using the following mixed-effects model:
\begin{equation}
y_{ij} = \beta_{j} x_{ij} + \mathbf{z}_i^\top \mathbf{\gamma}_{j} + u_i + \epsilon_{ij}, \qquad i = 1,\ldots, N, \quad j = 1,\ldots,K. \label{eq: mixed-effect}
\end{equation}
Here, $N$ is the number of subjects and $K$ is the number of experiments. 
For subject $i$ and experiment $j$, $y_{ij}$ is the response, $x_{ij} \in \{-1,1\}$ is the treatment/control assignment, and $\mathbf{z}_i$ is a $p$-dimensional covariate vector of whose first component is 1 (the intercept).  
The unknown parameters $\beta_j\in \mathbb R$ and $\mathbf{\gamma}_j\in\mathbb R^p$ encode, respectively, the treatment effect and the covariate effects in experiment $j$. 
The subject-specific random effects are denoted by $u_i$'s, which are
independent and identically distributed (i.i.d.) random variables with mean zero and variance $\tau^2$. 
The experimental errors are denoted by 
$\epsilon_{ij}$'s, which are i.i.d. random variables with mean zero and variance
$\sigma_j^2$.
We assume that random effects and experimental errors are independent.
Under this model, it follows that 
\[
\text{Cov}(y_{ij},y_{ij'}) = \text{Cov}(u_i + \epsilon_{ij}, u_i + \epsilon_{ij'}) = \text{Var}(u_i) = \tau^2,
\]
for any subject $i$ and two experiments $j$ and $j'$. 
This nonzero covariance characterizes the dependence between responses across experiments.

As is common in experimental design with subject covariates, we are primarily interested in the treatment effects rather than the covariate effects \citep{bhat2020near}.
Accordingly, we seek designs that maximize the precision of the treatment-effect estimates. 
Designing an experiment to maximize the precision of parameter estimates is known as $D_s$-optimality \citep{lim1988efficient}. 
In our setting, the parameters of interest are $\bm{\beta}^\top = [\beta_1, \beta_2, \ldots, \beta_K]$, and the corresponding estimates are $\hat{\bm{\beta}}^\top = [\hat{\beta}_1, \hat{\beta}_2,\ldots, \hat{\beta}_K]$. 
Assuming that $\tau^2$ and $\sigma_j^2$ are known for $j=1,\ldots, K$, the covariance structure implied by \eqref{eq: mixed-effect} yields the generalized least squares (GLS) estimator \citep{christensen2002plane} and its precision matrix. 
Let  
\[
\mathbf{Y} = [y_{1,1},\ldots,y_{N,1},y_{1,2},\ldots,y_{N,2},\ldots,y_{1,K},\ldots,y_{N,K}]^\top.
\]
be the stacked response vector. 
The GLS estimates of the treatment and covariate effects are given by
\begin{align} \label{eq: GLS-est}
\left(
\hat{\bm{\beta}}^\top,
\hat{\bm{\gamma}}^\top
\right)^\top = (X^\top V^{-1} X)^{-1} X^\top V^{-1} \mathbf{Y},
\end{align}
where $\hat{\bm{\beta}}$ and $\hat{\bm{\gamma}}$ are the estimated treatment and covariate effects in the experiments, and
\[
X = \left[
\operatorname{diag}(\mathbf{x}_1,\ldots,\mathbf{x}_K)
\;\middle|\;
I_K \otimes Z
\right]
\]
is the design matrix.
$\otimes$ is the Kronecker product.
$\mathbf{x}_j$ is the $N\times 1$ vector of allocations $x_{ij}$ for subjects $i = 1,\ldots,N$ in experiment $j$, and $Z$ is the $N\times p$ matrix with the $i$th row $\mathbf{z}_i^\top$. 
Throughout this paper, we assume that $Z$ is full rank.
In addition, the covariance matrix of the response vector $\mathbf Y$ is given by
$
V = \operatorname{diag}(\sigma_1^2,\ldots,\sigma_K^2)
    \otimes I_N
    + \tau^2 (\mathbf{1}_K\mathbf{1}_K^\top) \otimes I_N$,
where $\mathbf{1}_K$ is a $K$-dimensional vector of ones, and $I_N$ is an identity matrix of size $N\times N$.

The matrix $X^\top V^{-1} X$ in \eqref{eq: GLS-est} is the precision matrix of $(\hat{\bm{\beta}}^\top, \hat{\bm{\gamma}}^\top)^\top$. 
Block matrix inversion can then be applied to $X^\top V^{-1} X$ to obtain the covariance matrix of $\hat{\bm{\beta}}$, and hence its precision matrix. 
The resulting precision matrix for $\hat{\bm{\beta}}$ is stated in Proposition \ref{prop: precision-treat} (proof in the Supplementary Material).
\begin{proposition} \label{prop: precision-treat}
Under the model assumption in \eqref{eq: mixed-effect}, the precision of the treatment effect estimates $\hat{\bm{\beta}} = \{\hat{\beta}_j: j = 1,\ldots,K\}$ is given by 
\begin{align}\label{eq:p}
P(\hat{\bm{\beta}}) =\frac{1}{c}
\begin{bmatrix}
Q_1 \mathbf{x}_1^\top P_{Z^\perp} \mathbf{x}_1 & R_{1,2} \mathbf{x}_1^\top P_{Z^\perp} \mathbf{x}_2 & \ldots & R_{1,K} \mathbf{x}_1^\top P_{Z^\perp} \mathbf{x}_K\\
R_{2,1} \mathbf{x}_2^\top P_{Z^\perp} \mathbf{x}_1 & Q_2 \mathbf{x}_2^\top P_{Z^\perp} \mathbf{x}_2 & \ldots & R_{2,K} \mathbf{x}_2^\top P_{Z^\perp} \mathbf{x}_K\\
\vdots & \vdots & \ddots & \vdots \\
R_{K,1} \mathbf{x}_K^\top P_{Z^\perp} \mathbf{x}_1 & R_{K,2} \mathbf{x}_K^\top P_{Z^\perp} \mathbf{x}_2 & \ldots & Q_K \mathbf{x}_K^\top P_{Z^\perp} \mathbf{x}_K
\end{bmatrix},
\end{align}
where $P_{Z^\perp} = I - Z(Z^\top Z)^{-1} Z^\top$,
$c = 1 + \tau^2 \sum_{\ell = 1}^K \sigma_{\ell}^{-2}$, $Q_j = \sigma_j^{-2}(c - \tau^2\sigma_j^{-2})$, and $R_{j,j'} = -\tau^2 \sigma_j^{-2} \sigma_{j'}^{-2}$.
\end{proposition}

As mentioned earlier, our goal is to choose the design allocations $\mathbf{x}_1,\ldots,\mathbf{x}_K$ that maximize the determinant of the precision matrix in \eqref{eq:p}:
\begin{align} \label{prob: d-opt}
\underset{\mathbf{x}_1,\ldots,\mathbf{x}_K \in \{-1,1\}^N}{\text{max}} \ \left[\text{det}\Big(P(\hat{\bm{\beta}})\Big)\right]^{1/K},
\end{align}
where $\text{det}(\cdot)$ denotes the determinant. The $K$-th root 
of the determinant
normalizes the design criteria through the number of experiments.
As can be seen from the precision matrix in Proposition \ref{prop: precision-treat}, allocations that achieve covariate balance within experiments and orthogonality across experiments tend to have higher D-efficiency. 
Covariate balance increases the diagonal terms $\frac{1}{c} Q_j \mathbf{x}_j^\top P_{Z^\perp} \mathbf{x}_j$ while orthogonality (together with covariate balance) reduces the off-diagonal terms $\frac{1}{c} R_{j,j'} \mathbf{x}_j^\top P_{Z^\perp} \mathbf{x}_{j'}$ for $j \neq j'$. 
\begin{theorem} \label{thm:upper}
Under the model assumptions in \eqref{eq: mixed-effect},
\[
 \mathrm{det}\Big(P(\hat{\bm{\beta}})\Big)\leq \frac{N^K}{c^K}\prod^K_{j=1}Q_j.
\]
The upper bound is attained if the treatment allocation vectors are perfectly covariate balanced and mutually orthogonal, i.e.,
\begin{align}
Z^\top \mathbf{x}_j
    &= \bm{0},
    && j = 1,\ldots,K,
    \label{eq:covariate-balance}\\
\mathbf{x}_j^\top \mathbf{x}_{j'}
    &= 0,
    && j \neq j', \quad j,j' \in \{1,\ldots,K\}.
    \label{eq:orthogonal}
\end{align}
\end{theorem}

Theorem~\ref{thm:upper} identifies key requirements for the design of collaborative experiments. 
We next establish theoretical properties of the variance of an individual treatment-effect estimator to quantify the benefit of collaborative design and analysis. 
For the $k$-th experiment, the collaborative-analysis estimator $\hat\beta_k$ is the $k$th entry of \eqref{eq: GLS-est}, and its variance is
\[
\mathrm{Var}(\hat\beta_k)=c\left[Q_k \mathbf{x}_k^\top P_{Z^\perp} \mathbf{x}_k-\mathbf{x}^\top_k P_{Z^\perp} \bm R_{k,-k}\bm R^{-1}_{-k, -k}\bm R^\top_{k,-k} P_{Z^\perp} \mathbf{x}_k\right]^{-1}.
\]
Here
$\bm R_{k,-k}=[R_{k,1}\mathbf{x}_1 \ \ldots \ R_{k,k-1}\mathbf{x}_{k-1} \  R_{k,k+1}\mathbf{x}_{k+1} \ \ldots \ R_{k,K}\mathbf{x}_{K}]$
and $\bm R_{-k,-k}$ is a $(K-1)\times(K-1)$ matrix with diagonal entries
$\left(\bm R_{-k,-k}\right)_{j,j}=Q_j\mathbf{x}_j^\top P_{Z^\perp} \mathbf{x}_j$
and off-diagonal entries (for  $j\neq j'$)
$\left(\bm R_{-k,-k}\right)_{j,j'}=R_{j,j'}\mathbf{x}_j^\top P_{Z^\perp} \mathbf{x}_{j'}$.
For individual (non-collaborative) analysis, the estimation of $\beta_k$ is based only on outcomes from experiment $k$, $\mathbf{y}_k=[y_{1,k},\ldots,y_{N,k}]^\top$. 
For $k=1, \ldots, K$, let $\tilde{\beta}_k$ denote the least squares estimator of $\beta_k$. 
Then
\[
\tilde \beta_k=\frac{\mathbf{x}_k^\top P_{Z^\perp} \mathbf{y}_k}{\mathbf{x}_k^\top P_{Z^\perp} \mathbf{x}_k}\quad\mathrm{and}\quad
\mathrm{Var}(\tilde \beta_k)=\frac{\sigma^2_k+\tau^2}{\mathbf{x}_k^\top P_{Z^\perp} \mathbf{x}_k}.
\]
Since both $\hat\beta_k$ and $\tilde\beta_k$ are unbiased for $\beta_k$, we compare their efficiency via their variances.

\begin{theorem} \label{thm:var}
For $k=1, \ldots, K$, we have 
\[
\frac{c}{NQ_k}\leq \mathrm{Var}(\hat{\beta}_k)\leq \mathrm{Var}(\tilde{\beta}_k).
\]
The lower bound is attained if the allocation vector for experiment $k$ is perfectly covariate balanced and  orthogonal to all other allocation vectors, i.e.,
$
Z^\top \mathbf{x}_k = \bm{0}\quad\mathrm{and}\quad \mathbf{x}_j^\top \mathbf{x}_{k} = \bm{0} \quad \forall j \neq k$.
The upper bound is attained when all experiments use the same allocation vector, i.e., 
$
\mathbf{x}_1=\mathbf{x}_2=\cdots=\mathbf{x}_K
$.
\end{theorem}

\begin{figure}[htbp]
 \centering
 \includegraphics[scale=0.65]{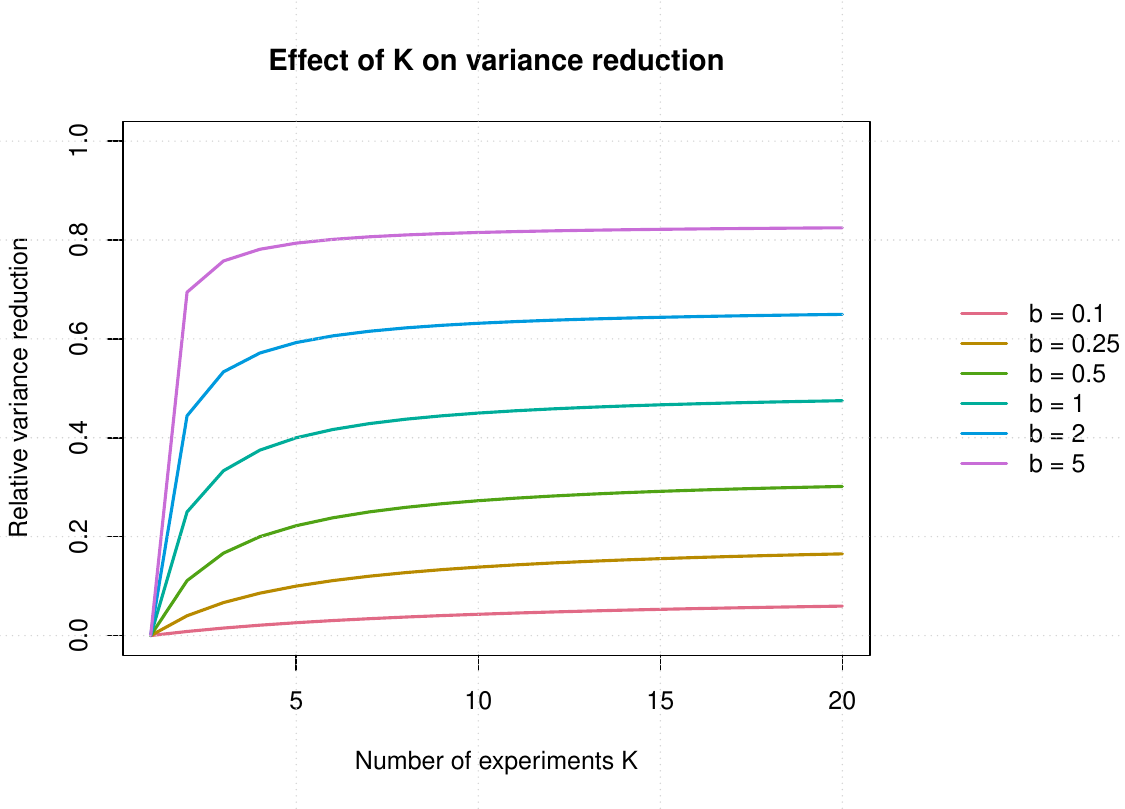}
 \caption{Relative variance reduction as the number of experiments $K$ increases, for different values of $b = \tau^2 / \sigma^2$.}
 \label{fig:relative-variance-reduction}
\end{figure}

The result from this theorem shows that collaborative analysis can always reduce the variance of a treatment-effect estimator relative to analyzing the corresponding experiment alone, regardless of allocation choices. 
When an experiment's allocation vector is perfectly covariate balanced and orthogonal to those of all other experiments, the variance is minimized, illustrating an additional benefit of collaborative design. 
In this ideal case, the relative variance reduction for experiment $k$ is 
\begin{equation}\label{eq:reduce}
\frac{\mathrm{Var}(\tilde{\beta}_k)-\mathrm{Var}(\hat{\beta}_k)}{\mathrm{Var}(\tilde{\beta}_k)}=
\frac{\tau^4\sigma^{-2}_k\sum^K_{j\neq k}\sigma^{-2}_j}{(1+\tau^2\sigma^{-2}_k)(1+\tau^2\sum^K_{j\neq k}\sigma^{-2}_j)}.
\end{equation}
This quantity equals zero if and only if $\tau^2=0$. 
The expression also shows that the benefit of collaborative design and analysis increases as the random-effect variance $\tau^2$ increases. 
Moreover, the reduction is largest when $\sigma_k^{-2}$ is close to $\sum_{j\neq k}^K \sigma_j^{-2}$, indicating that experiment $k$ is about as informative as the collection of all remaining experiments. 
To further simplify, assume $\sigma_1=\ldots=\sigma_K=\sigma$. 
Then \eqref{eq:reduce} becomes
\[
\frac{\mathrm{Var}(\tilde{\beta}_k)-\mathrm{Var}(\hat{\beta}_k)}{\mathrm{Var}(\tilde{\beta}_k)}=
\frac{b^2(K-1)}{(1+b)(1+b(K-1))},
\]
with $b=\tau^2/\sigma^2$. 
As shown in Figure~\ref{fig:relative-variance-reduction}, the relative variance reduction increases with $K$ and depends on $b$. 
However, as $K$ grows, the marginal gain in variance reduction diminishes. 
Overall, collaborative analysis consistently reduces variance, with greater reductions when the random-effect variance is larger relative to the noise variance.
From a practical perspective, collaborative analysis is most beneficial for groups of experiments with similar precision and substantial shared subject-level heterogeneity.

\section{Efficient Algorithms for Design Construction} \label{sec: algo}

In this section, we propose several efficient algorithms from two complementary angles to find design allocations that align with the D-optimality criterion in \eqref{prob: d-opt}. For notational convenience, we use the determinant of the precision matrix (without taking the $K$-th root) as the optimization objective in this section.

\subsection{Sequential-decomposition based Greedy Algorithms} \label{subsec: greedy}
We propose greedy algorithms for constructing designs for multiple controlled experiments. 
The main idea is straightforward: we decompose \eqref{prob: d-opt} into a sequence of $K$ subproblems. 
In subproblem $j$, we construct an allocation for experiment $j$ by solving a binary quadratic program derived from the block-determinant formula \citep{abadir2005matrix} and incorporating information from the previous $j-1$ allocations. 
We propose two approaches for solving the binary quadratic program in each subproblem: (1) semi-definite programming (SDP) relaxation based randomized algorithm \citep{ben2001lectures}; and (2) binary quadratic programming (BP) using deterministic integer programming techniques.


We now provide the technical details of the proposed algorithms. 
Let $P(\hat{\bm{\beta}})_{1:j-1}^{1:j-1}$ denote the submatrix of $P(\hat{\bm{\beta}})$ in \eqref{eq:p} formed by the first $j-1$ rows and columns.
Next, let $P(\hat{\bm{\beta}})_{1:j-1}^j$ denote the vector consisting of the first $j-1$ entries of the $j$th column of $P(\hat{\bm{\beta}})$, i.e., $P(\hat{\bm{\beta}})_{1:j-1}^j = \frac{1}{c} \ [R_{1,j}\mathbf{x}_1^\top P_{Z^\perp}\mathbf{x}_j, R_{2,j}\mathbf{x}_2^\top P_{Z^\perp}\mathbf{x}_j,\ldots,R_{j-1,j}\mathbf{x}_{j-1}^\top P_{Z^\perp}\mathbf{x}_j]^\top$.
Lastly, let $P(\hat{\bm{\beta}})_j^j$ denote the $j$th diagonal entry of $P(\hat{\bm{\beta}})$. 
Then, assuming that $P(\hat{\bm{\beta}})_{1:j-1}^{1:j-1}$ is invertible, and using the forms for $P(\hat{\bm{\beta}})_{1:j-1}^{1:j-1}$, $P(\hat{\bm{\beta}})_{1:j-1}^j$, and $P(\hat{\bm{\beta}})_j^j$, it follows from the formula for calculating the determinant of a block matrix that
\begin{align}\label{eq: det pbjj}
\text{det}\Big(P(\hat{\bm{\beta}})_{1:j}^{1:j}\Big) = \text{det}\Big( P(\hat{\bm{\beta}})_{1:j-1}^{1:j-1}\Big) \Big(P(\hat{\bm{\beta}})_j^j - (P(\hat{\bm{\beta}})_{1:j-1}^j)^\top \big(P(\hat{\bm{\beta}})_{1:j-1}^{1:j-1}\big)^{-1} P(\hat{\bm{\beta}})_{1:j-1}^j\Big).
\end{align}

Now, given allocations to the first $j-1$ experiments, denoted by $\mathbf{x}_1^*, \mathbf{x}_2^*,\ldots,\mathbf{x}_{j-1}^*$, the quantity $\det\big( P(\hat{\bm{\beta}})_{1:j-1}^{1:j-1}\big)$ is constant, and 
$
\Big(P(\hat{\bm{\beta}})_j^j - (P(\hat{\bm{\beta}})_{1:j-1}^j)^\top \big(P(\hat{\bm{\beta}})_{1:j-1}^{1:j-1}\big)^{-1} P(\hat{\bm{\beta}})_{1:j-1}^j\Big)
$
can be expressed as a quadratic form in $\mathbf{x}_j$. 
Specifically, this quadratic form is
\begin{align} \label{eq: quad}
q_j(\mathbf{x}_j|\mathbf x^\ast_1, \ldots, \mathbf x^\ast_{j-1}) =\mathbf{x}_j^\top \Big (\frac{1}{c}Q_j P_{Z^\perp} - B_{1:j-1} \big( P^*(\hat{\bm{\beta}})_{1:j-1}^{1:j-1}\big)^{-1} B_{1:j-1}^\top \Big) \mathbf{x}_j,
\end{align}
where $B_{1:j-1} = \frac{1}{c} \ \big[R_{1,j} P_{Z^\perp} \mathbf{x}_1^*,   R_{2,j} P_{Z^\perp} \mathbf{x}_2^*,  \ldots, R_{j-1,j}P_{Z^\perp} \mathbf{x}_{j-1}^*\big]$ is an $N \times (j-1)$ matrix, and $P^*(\hat{\bm{\beta}})_{1:j-1}^{1:j-1}$ is $P(\hat{\bm{\beta}})_{1:j-1}^{1:j-1}$ evaluated at the previously selected allocations $\mathbf{x}_1^*,\ldots,\mathbf{x}_{j-1}^*$. 
The problem of choosing the allocation $\mathbf{x}_j$ in \eqref{eq:sub} that yields the largest marginal improvement in \eqref{eq: det pbjj} given $\mathbf{x}_1^*, \mathbf{x}_2^*,\ldots,\mathbf{x}_{j-1}^*$ is equivalent to choosing $\mathbf{x}_j$ to maximize $q_j(\mathbf{x}_j|\mathbf x^\ast_1, \ldots, \mathbf x^\ast_{j-1})$ in \eqref{eq: quad}. 
Maximizing \eqref{eq: quad} for experiment $j$ requires balancing covariates within the $j$th experiment, which is captured by $\mathbf{x}_j^\top \big( \frac{1}{c}Q_j P_{Z^\perp}\big)\mathbf{x}_j$, and a form of orthogonality between experiment $j$ and pre-allocated experiments, captured by $-\mathbf{x}_j^\top \Big(B_{1:j-1} \big( P^*(\hat{\bm{\beta}})_{1:j-1}^{1:j-1}\big)^{-1} B_{1:j-1}^\top \Big)\mathbf{x}_j$. 
Accordingly, we approximate the original problem in \eqref{prob: d-opt} by the following sequence of subproblems:
\begin{align}\label{eq:sub}
\mathbf{x}_j^*
\in \underset{\mathbf{x}_j\in\{-1,1\}^N}{\operatorname{argmax}}
\,q_j(\mathbf{x}_j\mid\mathbf{x}_1^*,\ldots,\mathbf{x}_{j-1}^*),
\quad j=1,\ldots,K,
\end{align}
where $q_1(\mathbf{x}_1)=
\frac{Q_1}{c}\mathbf{x}_1^\top P_{Z^\perp}\mathbf{x}_1$.
Each subproblem can be solved using BP or the SDP relaxation based randomization approach, as noted earlier. 
The order of the experiments can be determined randomly or based on the practical timeline for conducting the experiments. In the simulation study, we will demonstrate that the ordering of experiments has a negligible effect on the resulting design performance.
Algorithm \ref{alg:greedy} summarizes the resulting procedure. 
We discuss BP and SDP for solving each subproblem in more details later. 

\begin{algorithm}
\caption{Greedy allocation via sequential decomposition}\label{alg:greedy}
\begin{algorithmic}
\Require $\tau, \sigma_1,\ldots,\sigma_K, Z$ 
\Ensure $\mathbf{x}_1^*,\mathbf{x}_2^*,\ldots,\mathbf{x}_K^*$
\State Compute $P_{Z^\perp}$ and $c$ (See Proposition \ref{prop: precision-treat}).
\For{$j \gets 1$ to $K$}
\State Compute $Q_j$ (See Proposition \ref{prop: precision-treat}).   
\If{$j>1$}
\State Compute $B_{1:j-1}$ and $P^*(\hat{\bm{\beta}})_{1:j-1}^{1:j-1}$ as discussed in Section \ref{subsec: greedy}.
\EndIf
\State Solve the $j$th problem in \eqref{eq:sub} via BP or SDP to obtain $\mathbf x^\ast_j$.
\EndFor
\end{algorithmic}
\end{algorithm}

\paragraph{BP with Deterministic Integer Programming Techniques.} For each subproblem in \eqref{eq:sub}, the binary quadratic program for allocation in experiment $j$ can be solved to optimality using deterministic integer programming techniques available in optimization solvers such as Gurobi \citep{gurobi}.
This approach leads to non-random allocation of subjects to treatment and control. 

\paragraph{SDP Relaxation based Randomization Algorithm.}
We next briefly introduce an SDP relaxation-based randomization algorithm that yields a randomized solution to \eqref{eq: quad} with a quality guarantee. 
This algorithm was originally proposed by \cite{goemans1995improved} for the Max-Cut problem \citep{commander2009maximum}, and was later generalized to solve general binary quadratic maximization problems \citep{ben2001lectures}. 
\cite{bhat2020near} proposed using  this randomized algorithm in their offline setting for designing a single experiment, citing its ability to provide a solution in polynomial time as an advantage over directly solving their original NP-hard problem. 
We use this algorithm because it yields efficient allocations while preserving randomness in the assignment. 

In our setting, the SDP relaxation-based randomized algorithm for finding an allocation for experiment $j$ works as follows. 
We first rewrite the objective in \eqref{eq: quad} and formulate the optimization problem as
\begin{align}\label{prob: trace}
 \underset{\mathbf{x}_j \in \{-1,1\}^N}{\text{max}}\text{Tr}\Bigg( \Big (\frac{1}{c}Q_j P_{Z^\perp} - B_{1:j-1} \big( P^*(\hat{\bm{\beta}})_{1:j-1}^{1:j-1}\big)^{-1} B_{1:j-1}^\top \Big) \mathbf{x}_j \mathbf{x}_j^\top \Bigg),
\end{align}
where $\text{Tr}(\cdot)$ denotes the trace of a matrix. 
The equivalence of the objectives in \eqref{eq: quad} and \eqref{prob: trace} follows from the cyclic property of the trace function. 
The matrix $\mathbf{x}_j \mathbf{x}_j^\top$ is positive semi-definite, with all diagonal elements equal to 1. 
Moreover, the matrix $\frac{1}{c}Q_j P_{Z^\perp} - B_{1:j-1} \big( P^*(\hat{\bm{\beta}})_{1:j-1}^{1:j-1}\big)^{-1} B_{1:j-1}^\top$ is positive semi-definite by Sylvester's criterion \citep{prussing1986principal}. 

A natural SDP relaxation of \eqref{prob: trace} is given by \begin{align}\label{prob: SDP}
\underset{W_j}{\text{max}} \ &\text{Tr}\Bigg( \Big (\frac{1}{c}Q_j P_{Z^\perp} - B_{1:j-1} \big( P^*(\hat{\bm{\beta}})_{1:j-1}^{1:j-1}\big)^{-1} B_{1:j-1}^\top \Big) W_j \Bigg)\\
    s.t.\quad & 
    (W_j)_{ii} = 1 , \ i = 1,\ldots,N \quad \text{and} \quad W_j \succeq \mathbf{0}_{N\times N} \nonumber.
\end{align}
The constraint $(W_j)_{ii} = 1$ enforces the diagonal entries of $W_j$ equal 1 and $W_j \succeq \mathbf{0}_{N\times N}$ enforces that $W_j$ is positive semi-definite. 
This SDP can be solved in optimization solvers such as MOSEK \citep{mosek}.

A randomized allocation for experiment $j$ that approximately solves \eqref{eq: quad} is constructed as follows. 
We solve \eqref{prob: SDP} to optimality to obtain $W_j^*$, and then find a matrix $(W_j^*)^{1/2}$ such that $W_j^* = \big((W_j^*)^{1/2}\big)^\top (W_j^*)^{1/2}$. 
This can be achieved via eigenvalue decomposition or a Cholesky factorization of a diagonally perturbed version of $W_j^*$ \citep{trefethen2022numerical}. 
Next, we generate a random vector $v_j$ uniformly on the unit $N$-sphere; this can be done by sampling $v_j' \sim N(\mathbf{0}_{N \times 1}, I_N)$ and setting $v_j=v_j'/\lVert v_j'\lVert_2$ \citep{muller1959note}. 
Subject $i$ is then assigned to treatment (i.e., $x_{ij}=1$) or control (i.e., $x_{ij}=-1$) in experiment $j$ using the following procedure for $i = 1,\ldots,N$:
\begin{align}\label{fun: assign}
x_{ij} = \begin{cases}
1 & \text{ if } v_j^\top ((W_j^*)^{1/2})_{1:N}^i \geq 0\\
-1 & \text{ if } v_j^\top ((W_j^*)^{1/2})_{1:N}^i < 0
\end{cases},
\end{align}
where $((W_j^*)^{1/2})_{1:N}^i$ is the $i$th column of $(W_j^*)^{1/2}$. 
Here, the $v_j$ are generated independently of each other. 

\subsection{Large-Scale Design based on Randomized Algorithm} \label{subsec: single}

The greedy algorithm proposed in Section \ref{subsec: greedy} requires solving a sequence of subproblems in \eqref{eq:sub}. 
When both the number of experiments and the number of subjects are large, which is common in industrial practice where hundreds of controlled experiments may be executed daily \citep{larsen2024statistical},
the randomized greedy algorithm may be prohibitively expensive to compute. 
To address this large-scale challenge, we develop a Single Design based Randomized algorithm (SDR) for design construction.
The idea of SDR is to solve a single SDP with $N^2$ decision variables and then use a randomization scheme to produce allocations for each experiment. 
The SDR addresses covariate balance within experiments, while the randomization scheme targets orthogonality across experiments. 
This contrasts with the randomized greedy algorithm in Section \ref{subsec: greedy}, which enforces both objectives through a sequence of SDPs.

The SDR is motivated by the observation that, when considered separately, the optimal allocation for each individual experiment is the same because each experiment's precision is represented by the same quadratic form, up to constant multiplicative factors. 
Specifically, the precisions of experiments $j$ and $j'$ are $\frac{1}{c}Q_j \mathbf{x}_j^\top P_{Z^\perp} \mathbf{x}_j$ and $\frac{1}{c}Q_{j'} \mathbf{x}_{j'}^\top P_{Z^\perp} \mathbf{x}_{j'}$, respectively; therefore, maximizing these two quadratic forms is equivalent. 
Thus, solving 
\begin{align}\label{eq: single-prec}
\underset{\mathbf{x}\in \{-1,1\}^N}{\text{max}} \ \mathbf{x}^\top P_{Z^\perp} \mathbf{x}
\end{align}
provides an allocation $\mathbf{x}$ that can be reused across experiments to achieve covariate balance within each experiment. 
However, using the same allocation $\mathbf{x}$ for every experiment reduces orthogonality across experiments and thus results in lower D-efficiency. 
We address this by solving an SDP relaxation of \eqref{eq: single-prec}
as in Section \ref{subsec: greedy}:
\begin{align}\label{eq: sdp}
    \underset{W}{\text{max}} \ &\text{Tr}\Big( P_{Z^\perp} W \Big)\\
    s.t.\quad & 
    (W)_{ii} = 1 , \ i = 1,\ldots,N \quad \text{and} \quad W \succeq \mathbf{0}_{N\times N} \nonumber.
\end{align}
Given the solution $W^*$ to the SDP above, we find $(W^*)^{1/2}$ 
such that $W^* = \big((W^*)^{1/2}\big)^\top (W^*)^{1/2}$.

Next, we use a randomly generated set of $K$ orthogonal vectors on the unit $N$-sphere to construct the allocation. To generate these orthogonal vectors, we first generate a vector $v_1$ uniformly from the unit $N$-sphere. We then utilize complete QR factorization \citep{trefethen2022numerical} to construct $N$ mutually orthogonal unit-length vectors $v_1,v_2,...,v_N$. For clarity, we utilize complete QR factorization to write $v_1$ in the following manner:
\begin{align}\label{eq:qr}
    v_1 = VR, \quad \text{where} \quad V = [v_1,v_2,...,v_N]\quad\mathrm{and}\quad R = [1,0,...,0]^T.
\end{align}
Here, $V$ is an $N\times N$ matrix where the columns are mutually orthogonal with unit length, and $R$ is a vector of size $N$.
We then uniformly select 
$K$ vectors from among these $N$ vectors coming from the columns of $V$, denoted by $\tilde{v}_1,\tilde{v}_2,\ldots,\tilde{v}_K$. The allocation of the $k$-th experiment is given by $\tilde v_k$ and $(W^*)^{1/2}$ using \eqref{fun: assign}.
The proposed algorithm is provided in Algorithm \ref{alg:large}.

\begin{algorithm}
\caption{Large-scale design based on randomized algorithm}\label{alg:large}
\begin{algorithmic}
\Require $Z$ 
\Ensure $\mathbf{x}_1^*,\mathbf{x}_2^*,\ldots,\mathbf{x}_K^*$

\State Solve the SDP problem in \eqref{eq: sdp}, and find $(W^*)^{1/2}$ satisfying $W^* = \big((W^*)^{1/2}\big)^\top (W^*)^{1/2}$
\State Generate mutually orthogonal and unit-length vectors $v_1,v_2,\ldots,v_N$ as in \eqref{eq:qr}.
\State Uniformly select $K$ vectors $\tilde{v}_1,\tilde{v}_2,\ldots,\tilde{v}_K$ from $v_1,\ldots,v_N$.
\For{$j \gets 1$ to $K$}
    \For{$i \gets 1$ to $N$}
        Calculate $x_{ij}$ according to (\ref{fun: assign}) using $\tilde{v}_j$ and $(W^*)^{1/2}$.
    \EndFor
\EndFor

\end{algorithmic}
\end{algorithm}

\section{Simulation Study}\label{sec: sim}
In this section, we investigate the D-efficiency of our proposed design methods, as well as the variances of the estimated treatment effects  associated with the first (i.e., $j=1$) experiment and last experiment (i.e., $j=K$) in the sequence in \eqref{eq:sub}.
Our investigation is for two purposes. First, it can examine the D-efficiency, an overall measure of efficiency across the full collection of experiments. We can also examine  the variances of individual experiments, which are local measures of efficiency.
Second, it is useful to examine whether the order of experiments affects the resulting variances when using our greedy methods, which allocate one experiment at a time. 
The scenarios comprise combinations of several factors, including the number of controlled experiments conducted, the amount of noise in the subject-specific error term, and the number of covariates. 

Here we describe the simulation setup in detail. 
First, we consider two values for the number of controlled experiments: $K\in \{4, 8\}$. 
Next, we consider a low- and a high-noise setting for the subject-specific error term, with $\tau=0.25$ and $\tau=2$, respectively. 
Across all scenarios, we set the experimental error variance to 1; that is, $\sigma_1 = \sigma_2 = \ldots = \sigma_K = 1$ for $K =4$ and 8. 
Lastly, we fix the number of subjects at 96, and consider $p\in \{10, 70\}$. 
The following are different design construction methods we compare in the simulation study.
\begin{enumerate}
\item (\textbf{RAND}): Fully random allocation; each subject is independently assigned to treatment or control in each experiment with probability $1/2$.
\item (\textbf{PB}): Plackett--Burman design \citep{plackett1946design}, which enforces balance within experiments and orthogonality across experiments. 
\item (\textbf{SDR}): Single design based randomized algorithm in Section \ref{subsec: single}.
\item (\textbf{GREED-SDP}): Greedy algorithm with each subproblem solved via an SDP relaxation in Section \ref{subsec: greedy}.
\item (\textbf{GREED-BP}): Greedy algorithm with each subproblem solved via deterministic binary programming in Section \ref{subsec: greedy}.
\end{enumerate}

We now discuss the replication structure of the simulation study. 
For each choice of the number of covariates, $p\in\{10, 70\}$, we randomly generate five $96 \times p$ covariate matrices $Z_1, Z_2, \ldots, Z_5$, where the first column is all 1s and each row (apart from the fixed entry in the first column) is independently generated from a $(p-1)$-dimensional multivariate normal distribution with mean $0$ and covariance matrix $I_{p-1}$. 
We use these covariate matrices across all scenarios with the same value of $p$ to assess whether the methods' performance is stable across different covariates realization. 
We then specify the number of controlled experiments and the noise level in the subject-specific error term. 
For each pre-generated covariate matrix $Z_\ell$, we replicate  RAND, PB, SDR, and GREED-SDP 100 times to construct designs; we run GREED-BP only once since it is deterministic. 
For methods using SDPs or binary programming, we cap solution time at 50 seconds. 
For each method and replication, we record the D-efficiency of the resulting design (measured by the objective in \eqref{prob: d-opt}) and the variances of the first and last experiments, obtained by inverting the precision matrix and extracting the $(1,1)$ and $(K, K)$ entries, respectively.

\begin{figure}[p]
    \centering
    \includegraphics[width=\textwidth]{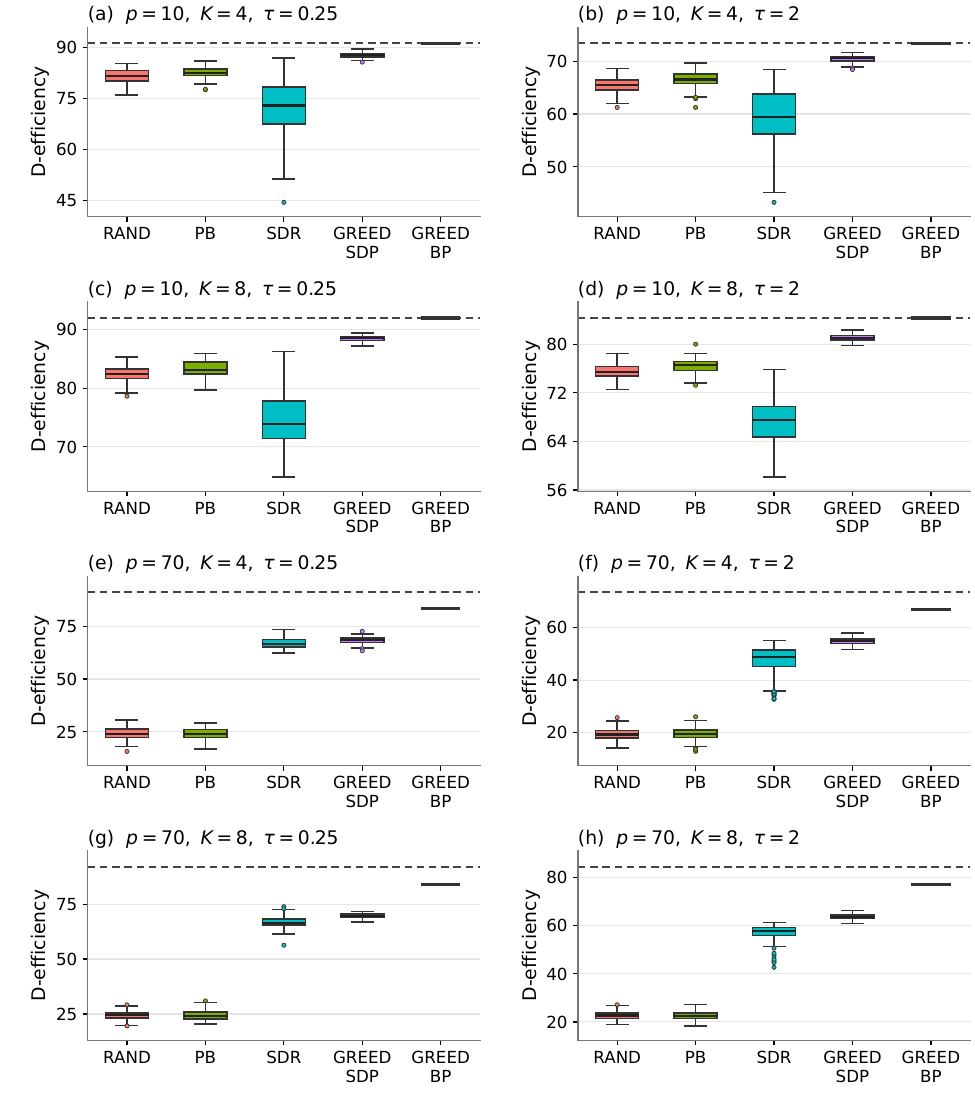}
    \caption{D-efficiencies of the methods RAND, PB, SDR, GREED-SDP, and GREED-BP across different experimental settings for covariate matrix $Z_1$. The dark gray dashed line is the upper bound in Theorem \ref{thm:upper}.}
    \label{fig:D-efficiencies}
\end{figure}

From the simulation study results, it is seen that that a key factor driving in method performance is the number of covariates. 
As the number of covariates increases from 10 to 70, the D-efficiency of covariate-agnostic methods such as RAND and PB decreases (Figure \ref{fig:D-efficiencies}). 
This is consistent with results on random, covariate-agnostic allocation in the presence of many covariates  \citep{bhat2020near}. 
However, when the number of covariates is small, covariate-agnostic methods may outperform our proposed SDR method. 
In addition, results were similar across different realizations of the covariates, suggesting that the methods have stable D-efficiency across covariate draws. 
The greedy methods GREED-SDP and GREED-BP typically perform the best: GREED-SDP performs on par with SDR in some cases in the 70-covariate regime, and GREED-BP always outperforms the other methods. 
We reiterate that although GREED-BP provides the highest D-efficiency, it does not provide randomized assignment and thus cannot mitigate bias from unmeasured covariates.  

\begin{figure}[p]
    \centering
    \includegraphics[width=\textwidth]{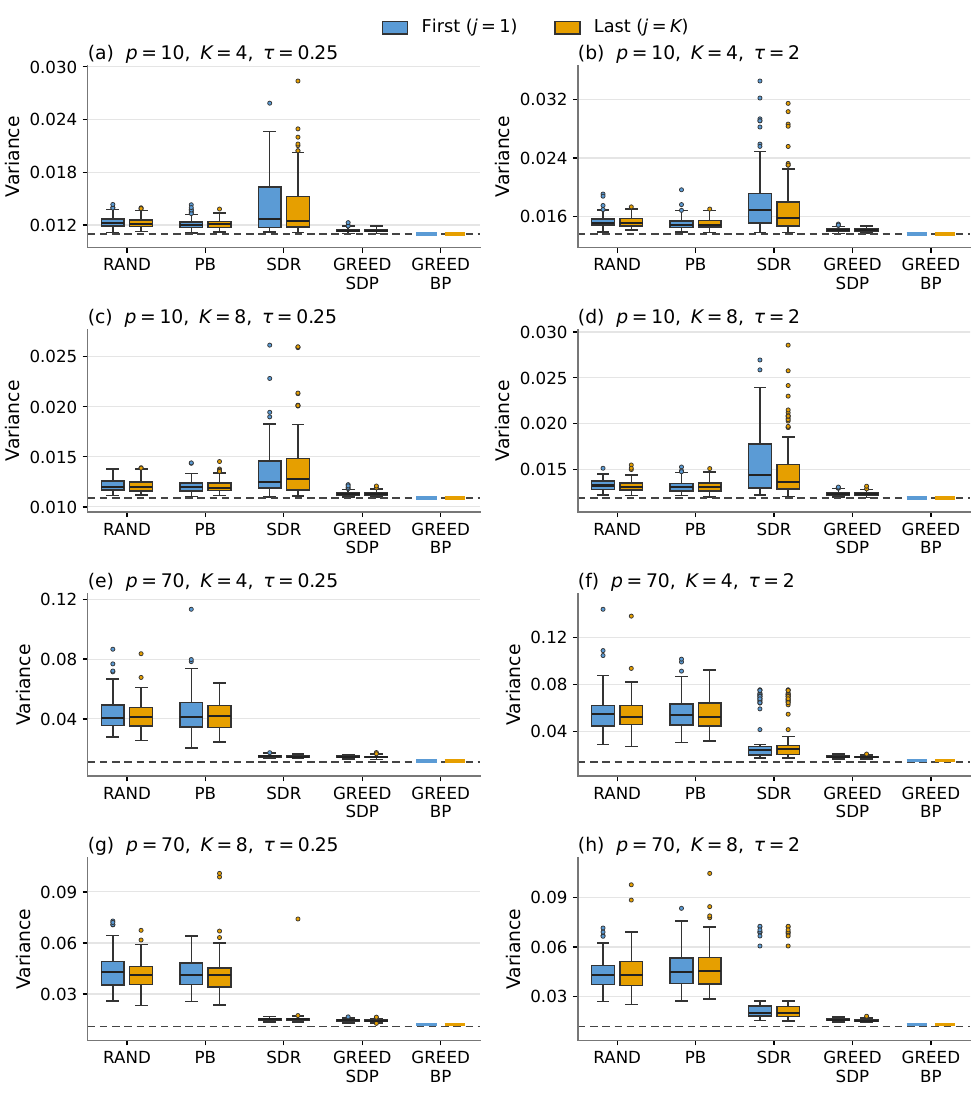}
    \caption{Variances of the first ($j=1$, blue) and last ($j=K$, orange) experiments of the methods RAND, PB, SDR, GREED-SDP, and GREED-BP across different experimental settings for covariate matrix $Z_1$. The dark gray dashed line is a theoretical lower bound of the variances.}
    \label{fig:variances}
\end{figure}


GREED-BP nearly achieves the upper bound in Theorem \ref{thm:upper} for D-efficiency in all cases with 10 covariates, suggesting that the greedy method performs well at producing covariate-balanced and orthogonal treatment allocations when the number of covariates is small relative to the number of subjects. 
In the 70-covariate setting, the gap between GREED-BP and the upper bound is non-negligible. 
In terms of reducing the variance of treatment effect estimates in individual experiments, there is not a large difference in the variances of treatment effect estimates between the first and last experiments, as shown in Figure \ref{fig:variances}. 
Similar to the case of D-efficiency, we see that our proposed methods outperform covariate-agnostic methods when the number of covariates is large. 
It is worth mentioning that while our GREED-BP method was unable to achieve the theoretical upper bound for D-efficiency, it was able to get close to the theoretical lower bound for the variance in the first and last experiment. 

\section{Case Study of Online Education Experiments}\label{sec: case}


We present a simulation-based case study motivated by ASSISTments \citep{heffernan2014assistments}, an online platform for elementary, middle, and high school mathematics.
ASSISTments also supports educational experiments, typically randomized controlled trials embedded in homework assignments.
Students entering an assignment are randomly assigned, with equal probability, to an experimental condition for the duration of that assignment.
In the assignments considered here, completion requires answering three consecutive questions correctly. This case study is also used for paired experiments in \cite{zhang2024collaborative}.

Two features of this setting motivate our collaborative design framework: subject covariates are available, and students may participate in multiple experiments through different homework assignments.
The recorded outcomes, assignment completion and the number of questions attempted, are binary and count responses, respectively.
Our design criterion is derived for continuous responses under the linear mixed-effects model in (\ref{eq: mixed-effect}), with variance parameters specified at the design stage.
We therefore evaluate the proposed continuous-response design methods using the observed covariates and shared-subject structure of ASSISTments.
The binary completion outcomes are used to calibrate a continuous-response simulator, allowing alternative treatment allocations to be compared in a setting informed by real data.
This case study evaluates design performance under the calibrated simulator.

\subsection{Setup}\label{subsec: casestudysetup}

The data comprise 22 randomized controlled trials described in \cite{selent2016assistments}.
We selected $K=3$ experiments conducted in assignments on ``Finding Expected Value,'' ``Conditional Probability,'' and ``Permutations and Combinations.''
To match our framework's assumption that every subject participates in every experiment, we restricted the data to the $N=193$ subjects who participated in all three experiments and had complete records for the covariates described below.

We use three recorded variables: ``average prior percent completion,'' ``average prior percent correct,'' and ``guessed gender.''
The first two measures vary across experiments. 
We average each over the three experiments for each subject to obtain the covariates $z_i^{(1)}$ and $z_i^{(2)}$.
Guessed gender has three recorded categories, ``male,'' ``female,'' and ``unknown,'' which we represent by two indicators: $z_i^{(3)}$ for ``male'' and $z_i^{(4)}$ for ``unknown,'' with ``female'' as the reference category.
The design covariate matrix includes an intercept and these four covariate columns.
Following the definition of $p$ in (\ref{eq: mixed-effect}), we consider total dimensions $p\in\{5,25,55,105\}$, including the intercept.
For $p>5$, we append an $N\times(p-5)$ matrix of simulated covariates, denoted by $\omega_i^{(1)},\ldots,\omega_i^{(p-5)}$, whose rows are independently drawn from $N(0,I_{p-5})$.
Thus, the four settings add 0, 20, 50, and 100 covariates, respectively.

For response replication $b$ and subject-effect scenario $s\in\{A,B\}$, the continuous-response simulator is
\begin{equation}\label{eq: pseudo-model}
    y_{ij}^{(b,s)} = \hat{\alpha}_j + \hat{\beta}_j x_{ij}
    + \sum_{r=1}^{4}\hat{\gamma}_j^{(r)}z_i^{(r)} + u_i^{(b,s)}
    + \theta\sum_{\ell=1}^{p-5}\omega_i^{(\ell)} + \epsilon_{ij}^{(b)},
\end{equation}
for $i=1,\ldots,N$ and $j=1,\ldots,K$.
Here, $x_{ij}\in\{-1,1\}$ is the treatment allocation supplied by the design being evaluated, and each added covariate has the common coefficient $\theta$.
The added covariate term is omitted when $p=5$.
The errors $\epsilon_{ij}^{(b)}$ are independently generated from $N(0,\sigma^2)$ across subjects, experiments, and replications.
The fitted coefficients determine the simulator's mean structure on the probit scale.
For calibration only, we fit a probit mixed-effects model to the observed completion outcomes, conditional on the recorded treatment assignments and covariates:
\begin{equation}\label{eq: glmm}
    \Phi^{-1}\!\left\{\Pr(\mathrm{Complete}_{ij}=1\mid u_i)\right\}
    = \alpha_j + \beta_j x_{ij} + \sum_{r=1}^{4}\gamma_j^{(r)}z_i^{(r)} + u_i,
\end{equation}
where $\Phi$ is the standard normal cumulative distribution function, and $u_i\overset{\mathrm{iid}}{\sim}N(0,\tau_{\mathrm{cal}}^2)$ are random intercepts independent of the covariates.
Responses are conditionally independent given these effects and the predictors.
We use the \texttt{glmer} function in \texttt{lme4} \citep{bates2015fitting} to obtain $\hat{\alpha}_j$, $\hat{\beta}_j$, $\hat{\gamma}_j^{(r)}$, and the fitted conditional modes $\tilde u_i$ of the subject effects.
Under the $\{-1,1\}$ coding, $2\beta_j$ is the treatment--control contrast on the probit scale.

The two subject-effect scenarios use
$u_i^{(b,A)}=\tilde u_i$ and $u_i^{(b,B)}\overset{\mathrm{iid}}{\sim}N(0,s_u^2)$, where 
$s_u\approx1.6733$ is the empirical standard deviation of the fitted conditional modes $\tilde u_i$, rather than the random-effect standard deviation estimated by the probit model.
In both scenarios, each subject's effect is shared across the three experiments.
Scenario A holds the fitted effects fixed across replications.
It therefore provides a comparison conditional on the fitted effects, which need not satisfy the normality and independence assumptions of the analysis model.
Scenario B generates new effects independently across subjects and replications, independently of the covariates and experimental errors.
It matches the empirical scale of the fitted effects but not their empirical distribution or association with the observed covariates.

We first compare D-efficiencies under subject-effect standard deviations $\tau\in\{0.1,1,10\}$ and common experimental-error standard deviations $\sigma\in\{0.1,1\}$.
For each covariate setting, we compare the original ASSISTments allocation, denoted ORIGINAL, with the five construction methods introduced in Section \ref{sec: sim}.
For $N=193$, the PB allocations are approximately balanced and orthogonal: treatment and control group sizes differ by one, and pairwise allocation inner products have magnitude one.
Each randomized method is replicated 30 times. 
GREED-BP is deterministic and is run once per setting, while ORIGINAL is the fixed allocation recorded in the data.
These comparisons use the specified variance parameters in the design criterion and do not require simulated responses.

We next compare the $L_2$ estimation errors of the treatment coefficients for the five constructed methods, RAND, PB, SDR, GREED-SDP, and GREED-BP, using responses from (\ref{eq: pseudo-model}).
For this comparison, GREED-BP and GREED-SDP designs are constructed using $\sigma=1$ and $\tau=s_u$; RAND, PB, and SDR retain their $\tau$-independent allocations.
The design value $\tau=s_u$ matches the generating subject-effect standard deviation in Scenario B and the empirical scale of the fixed effects in Scenario A.
Both scenarios use the same designs, and we set $\sigma=\theta=1$ in the response simulator.

For each method and covariate setting, we use one saved design: the designs and added covariates are generated separately from those used in the D-efficiency comparison.
Each design and its covariates are held fixed over 100 response replications per scenario.
Within each replication, all methods and both scenarios share the same experimental errors for each subject in each experiment.
The Scenario B subject-effect draws are also shared across methods and covariate settings.

For each simulated dataset, we fit a linear mixed-effects model by maximum likelihood, with a subject-specific random intercept and experiment-specific intercepts, treatment coefficients, and covariate coefficients.
The model includes all added covariates, and its variance components are estimated from the simulated responses.
Let $\bm{\beta}^{\star}=(\hat{\beta}_1,\hat{\beta}_2,\hat{\beta}_3)^\top$ denote the treatment-coefficient vector used to generate responses in both scenarios, calibrated from (\ref{eq: glmm}), and let $\widehat{\bm{\beta}}_m^{(b,s)}$ denote its estimate under method $m$ in replication $b$ of scenario $s$.
The $L_2$ estimation error is
\begin{equation}\label{eq: case_l2_error}
    L_{2,m}^{(b,s)} = \left\|\widehat{\bm{\beta}}_m^{(b,s)}-\bm{\beta}^{\star}\right\|_2
    = \left\{\sum_{j=1}^{3}\left(\widehat{\beta}_{m,j}^{(b,s)}-\beta_j^{\star}\right)^2\right\}^{1/2},
    \qquad b=1,\ldots,100.
\end{equation}
The resulting $L_2$ estimation errors are summarized using boxplots in the following subsection.

\subsection{Results}\label{subsec: casestudyresults}

We first summarize the D-efficiency results for $\sigma=1$. 
The results for $\sigma=0.1$ can be found in the Supplementary Material.
Figure \ref{fig:case_deff_combined} compares ORIGINAL and the five constructed designs across the four covariate settings and three values of $\tau$.
Each boxplot summarizes 30 allocations from a randomized method, while horizontal segments represent the single GREED-BP and ORIGINAL values.

\begin{figure}[p]
    \centering
    \captionsetup{font=small}
    \includegraphics[width=\textwidth]{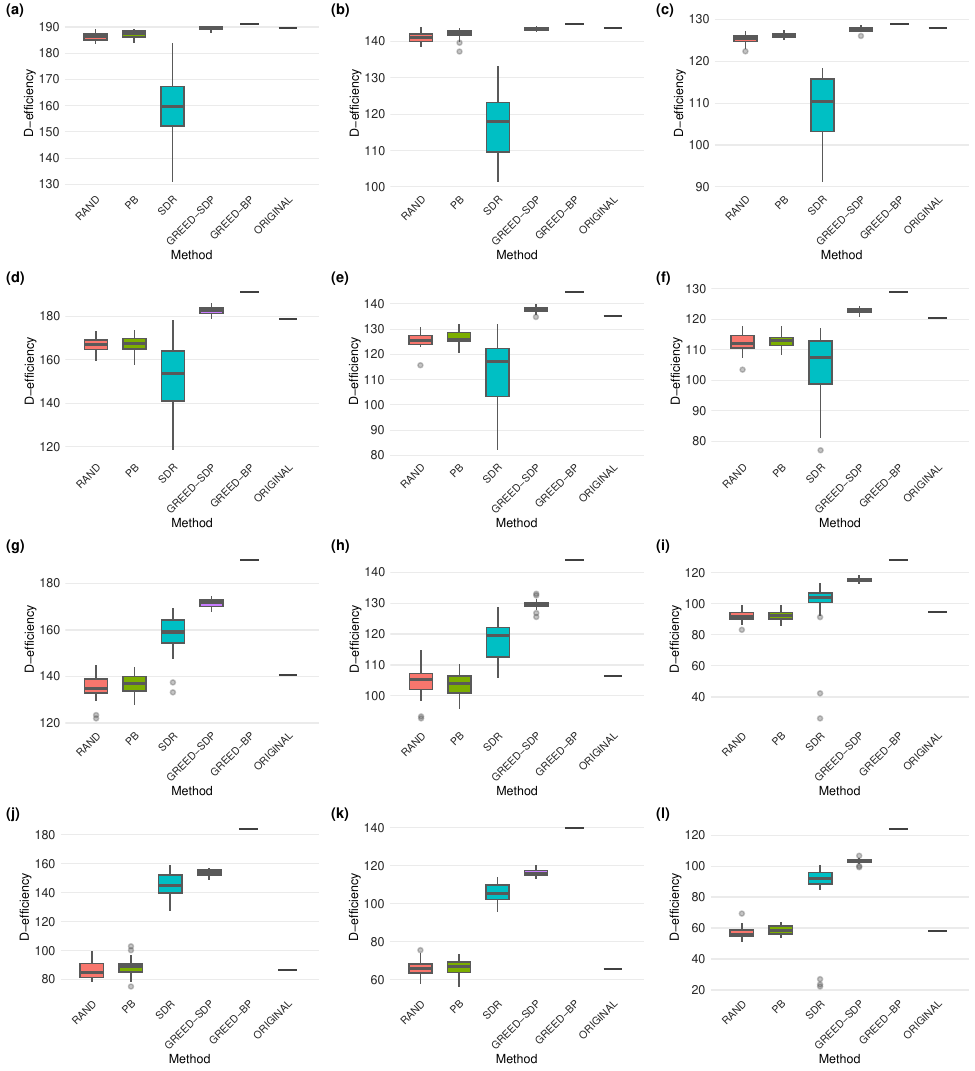}
    \caption{D-efficiencies for $\sigma=1$. Rows, from top to bottom, use total covariate dimensions $p=5,25,55,105$, including the intercept; columns, from left to right, use $\tau=0.1,1,10$. Panels (a)--(l) are ordered across rows. Boxplots summarize 30 designs per randomized method; horizontal segments show the single GREED-BP and ORIGINAL values.}
    \label{fig:case_deff_combined}
\end{figure}

In these comparisons, the relative advantage of the greedy methods increases with $p$, consistent with Section \ref{sec: sim}.
When $p=5$, ORIGINAL, RAND, PB, GREED-SDP, and GREED-BP have similar D-efficiencies, whereas SDR generally has lower and more variable efficiency.
With only four recorded covariate columns and $N$ subjects, explicitly optimizing covariate balance yields relatively modest gains over the covariate-agnostic allocations.

As $p$ increases to $25$, $55$, and $105$, GREED-BP and GREED-SDP show larger gains over ORIGINAL, RAND, and PB.
GREED-BP has the highest observed D-efficiency. 
For $p=25,55,105$, GREED-SDP ranks second by mean D-efficiency.
These results support GREED-SDP as an alternative that combines high D-efficiency with randomized allocation.
At $p=55,105$, SDR exceeds RAND and PB in mean D-efficiency but is more variable than GREED-SDP.
Unlike the sequential greedy methods, SDR solves a single covariate-balancing SDP and uses randomized rounding based on orthogonal vectors to encourage separation between experimental allocations; this procedure does not guarantee orthogonality of the resulting binary allocations.
ORIGINAL performs competitively in the low-dimensional setting but remains close to RAND and PB when many simulated covariates are added, consistent with its allocation having been generated without reference to the augmented covariate matrix.

One can see that these results agree with the precision-matrix structure in Proposition \ref{prop: precision-treat}.
For fixed variance parameters, covariate balance increases the diagonal contributions through $\mathbf{x}_j^\top P_{Z^\perp}\mathbf{x}_j$, while orthogonality of the projected allocations eliminates cross-experiment contributions through $\mathbf{x}_j^\top P_{Z^\perp}\mathbf{x}_{j'}$.
RAND and PB do not explicitly optimize these quantities.
In particular, $\mathbf{x}_j^\top\mathbf{x}_{j'}=0$ does not generally imply $\mathbf{x}_j^\top P_{Z^\perp}\mathbf{x}_{j'}=0$.
The greedy methods account for both covariate balance and previously selected allocations.
Although the D-efficiency scale changes with $\tau$, the broad performance pattern persists across the three values considered.

\begin{figure}[!h]
    \centering
    \includegraphics[scale=0.8]{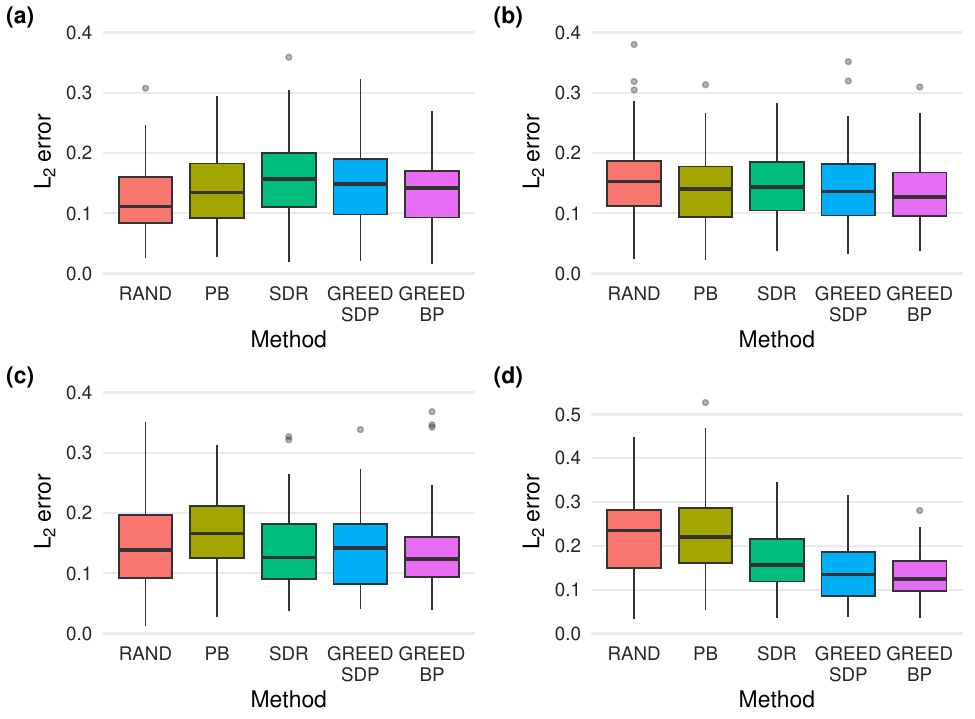}
    \caption{Scenario A: $L_2$ errors with fitted subject effects $u_i^{(b,A)}=\tilde{u}_i$ held fixed.
    Panels (a)--(d) correspond to $p=5,25,55,105$.
    }
    \label{fig:case_l2_error}
\end{figure}

\begin{figure}[!h]
    \centering
    \includegraphics[scale=0.8]{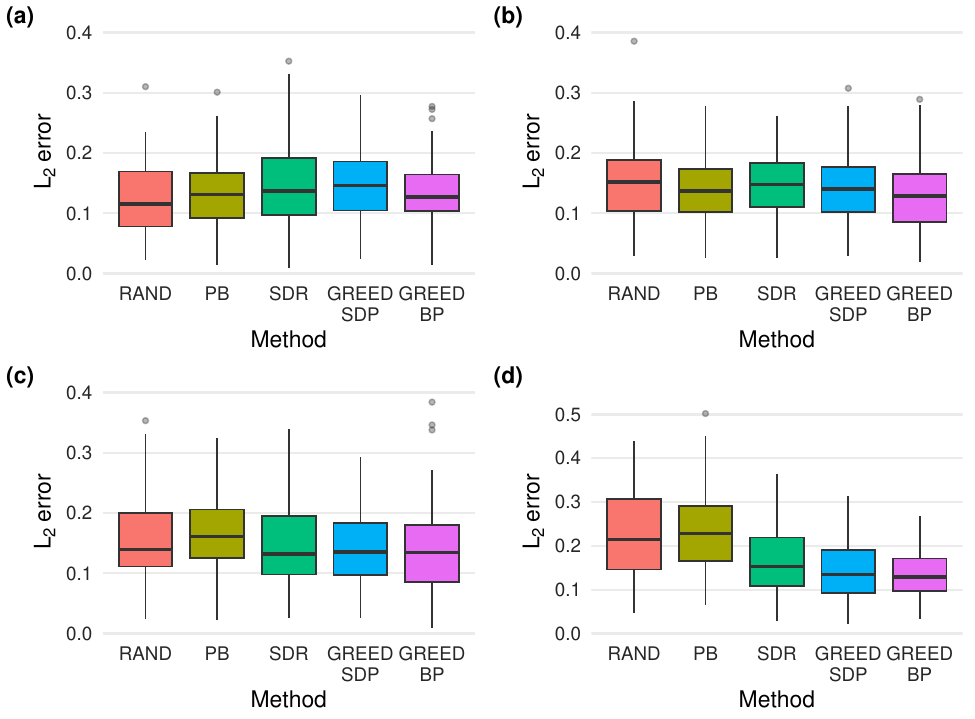}
    \caption{Scenario B: $L_2$ errors with subject effects independently redrawn from $N(0,s_u^2)$ in each replication.
    Panels (a)--(d) correspond to $p=5,25,55,105$, respectively.
    }
    \label{fig:case_l2_error_redrawn}
\end{figure}

Figures \ref{fig:case_l2_error} and \ref{fig:case_l2_error_redrawn} present the $L_2$ estimation errors for Scenarios A and B, respectively.
When $p=5$, the error distributions overlap substantially and RAND has the lowest mean error in both scenarios.
For $p=25$ and $p=55$, the distributions continue to overlap.
In both scenarios, all three covariate-aware methods have lower mean errors than RAND at $p=25$, but only GREED-BP has a lower mean error than PB.
At $p=55$, all three covariate-aware methods have lower mean errors than both RAND and PB.
The clearest separation occurs at $p=105$: RAND and PB have the largest mean errors, SDR is intermediate, and GREED-SDP and GREED-BP have the smallest mean and median errors in both scenarios.
For $p=105$, the mean $L_2$ errors for GREED-BP and GREED-SDP are 0.1303 and 0.1405, respectively, in Scenario A, and 0.1326 and 0.1423 in Scenario B.
Relative to RAND, these values correspond to reductions of approximately 42.7\% and 38.3\% in Scenario A, and 41.0\% and 36.6\% in Scenario B.

Thus, in the highest-dimensional setting, methods achieving higher D-efficiency also tend to yield more accurate treatment-effect estimates in both scenarios.
Higher D-efficiency does not necessarily imply a smaller realized $L_2$ error: D-efficiency is based on the determinant of the treatment-effect precision matrix, whereas the $L_2$ error also reflects response noise, potential bias, and variance-component estimation.
Nevertheless, for the saved designs and covariates considered here, the overall pattern supports the role of collaborative covariate-aware design in reducing treatment-effect estimation error, especially when many covariates are included.

\section{Conclusion}\label{sec: conc}

This work considered the collaborative design of multiple controlled experiments with shared subjects and observed subject covariates.
Using a linear mixed-effects model with subject-specific random effects, we derived the precision matrix of the treatment-effect estimators and established bounds on joint precision and individual treatment-effect variances.
We demonstrated that collaborative design can lead to  more precise estimates of treatment effects than the traditional method of individually designing and analyzing the experiments.
Two efficient approaches are proposed for constructing designs under the D-efficiency criterion.
The simulation study showed that the greedy methods generally achieved higher D-efficiency and smaller individual treatment-effect variances than covariate-agnostic methods when the covariate dimension was large relative to the number of subjects.
A continuous-response simulation calibrated using ASSISTments data showed a similar pattern, with the greedy methods yielding smaller treatment-effect estimation errors in the highest-dimensional setting.
These findings support the benefits of coordinating treatment allocations across experiments while accounting for subject covariates.


\nocite{higham2002accuracy,hadamard1893resolution}
\bibliography{ref}

@book{abadir2005matrix,
  title={Matrix algebra},
  author={Abadir, Karim M and Magnus, Jan R},
  volume={1},
  year={2005},
  publisher={Cambridge University Press}
}

@article{hadamard1893resolution,
  title={Resolution d'une question relative aux determinants},
  author={Hadamard, Jacques},
  journal={Bull. des Sciences Math.},
  volume={17},
  number={1},
  pages={240--246},
  year={1893}
}

@article{bhat2020near,
  title={Near-optimal ab testing},
  author={Bhat, Nikhil and Farias, Vivek F and Moallemi, Ciamac C and Sinha, Deeksha},
  journal={Management Science},
  volume={66},
  number={10},
  pages={4477--4495},
  year={2020},
  publisher={INFORMS}
}

@Article{bates2015fitting,
    title = {Fitting Linear Mixed-Effects Models Using {lme4}},
    author = {Douglas Bates and Martin M{\"a}chler and Ben Bolker and
      Steve Walker},
    journal = {Journal of Statistical Software},
    year = {2015},
    volume = {67},
    number = {1},
    pages = {1--48},
    doi = {10.18637/jss.v067.i01},
  }

@book{ben2001lectures,
  title={Lectures on modern convex optimization: analysis, algorithms, and engineering applications},
  author={Ben-Tal, Aharon and Nemirovski, Arkadi},
  year={2001},
  publisher={SIAM}
}

@article{bertsimas2015power,
  title={The power of optimization over randomization in designing experiments involving small samples},
  author={Bertsimas, Dimitris and Johnson, Mac and Kallus, Nathan},
  journal={Operations Research},
  volume={63},
  number={4},
  pages={868--876},
  year={2015},
  publisher={INFORMS}
}

@book{boyd2004convex,
  title={Convex optimization},
  author={Boyd, Stephen P and Vandenberghe, Lieven},
  year={2004},
  publisher={Cambridge university press}
}

@book{christensen2002plane,
  title={Plane answers to complex questions: the theory of linear models},
  author={Christensen, Ronald},
  year={2002},
  publisher={Springer}
}

@article{commander2009maximum,
  title={Maximum cut problem, MAX-cut.},
  author={Commander, Clayton W},
  journal={Encyclopedia of Optimization},
  volume={2},
  year={2009}
}

@article{goemans1995improved,
  title={Improved approximation algorithms for maximum cut and satisfiability problems using semidefinite programming},
  author={Goemans, Michel X and Williamson, David P},
  journal={Journal of the ACM (JACM)},
  volume={42},
  number={6},
  pages={1115--1145},
  year={1995},
  publisher={ACM New York, NY, USA}
}

@misc{gurobi,
  author = {{Gurobi Optimization, LLC}},
  title = {{Gurobi Optimizer Reference Manual}},
  year = 2023,
  url = "https://www.gurobi.com"
}

@article{heffernan2014assistments,
  title={The ASSISTments ecosystem: Building a platform that brings scientists and teachers together for minimally invasive research on human learning and teaching},
  author={Heffernan, Neil T and Heffernan, Cristina Lindquist},
  journal={International Journal of Artificial Intelligence in Education},
  volume={24},
  pages={470--497},
  year={2014},
  publisher={Springer}
}

@book{higham2002accuracy,
  title={Accuracy and stability of numerical algorithms},
  author={Higham, Nicholas J},
  year={2002},
  publisher={SIAM}
}

@article{imai2014covariate,
  title={Covariate balancing propensity score},
  author={Imai, Kosuke and Ratkovic, Marc},
  journal={Journal of the Royal Statistical Society Series B: Statistical Methodology},
  volume={76},
  number={1},
  pages={243--263},
  year={2014},
  publisher={Oxford University Press}
}

@article{koning2022experimentation,
  title={Experimentation and start-up performance: Evidence from A/B testing},
  author={Koning, Rembrand and Hasan, Sharique and Chatterji, Aaron},
  journal={Management Science},
  volume={68},
  number={9},
  pages={6434--6453},
  year={2022},
  publisher={INFORMS}
}

@article{larsen2024statistical,
  title={Statistical challenges in online controlled experiments: A review of a/b testing methodology},
  author={Larsen, Nicholas and Stallrich, Jonathan and Sengupta, Srijan and Deng, Alex and Kohavi, Ron and Stevens, Nathaniel T},
  journal={The American Statistician},
  volume={78},
  number={2},
  pages={135--149},
  year={2024},
  publisher={Taylor \& Francis}
}

@article{li2018balancing,
  title={Balancing covariates via propensity score weighting},
  author={Li, Fan and Morgan, Kari Lock and Zaslavsky, Alan M},
  journal={Journal of the American Statistical Association},
  volume={113},
  number={521},
  pages={390--400},
  year={2018},
  publisher={Taylor \& Francis}
}

@article{li2023optimal,
  title={Optimal Design of Controlled Experiments for Personalized Decision Making in the Presence of Observational Covariates},
  author={Li, Yezhuo and Zhang, Qiong and Khademi, Amin and Yang, Boshi},
  journal={The New England Journal of Statistics in Data Science},
  year={2023}
}

@article{lim1988efficient,
  title={Efficient $ D\_s $-Optimal Designs for Multivariate Polynomial Regression on the $ q $-Cube},
  author={Lim, Yong B and Studden, WJ},
  journal={The Annals of Statistics},
  volume={16},
  number={3},
  pages={1225--1240},
  year={1988},
  publisher={Institute of Mathematical Statistics}
}

@article{lyngbakken2020pragmatic,
  title={A pragmatic randomized controlled trial reports lack of efficacy of hydroxychloroquine on coronavirus disease 2019 viral kinetics},
  author={Lyngbakken, Magnus Nakrem and Berdal, Jan-Erik and Eskesen, Arne and Kvale, Dag and Olsen, Inge Christoffer and Rueegg, Corina Silvia and Rangberg, Anbj{\o}rg and Jonassen, Christine Monceyron and Omland, Torbj{\o}rn and R{\o}sj{\o}, Helge and others},
  journal={Nature communications},
  volume={11},
  number={1},
  pages={5284},
  year={2020},
  publisher={Nature Publishing Group UK London}
}

@article{morgan2012rerandomization,
  title={Rerandomization to improve covariate balance in experiments},
  author={Morgan, Kari Lock and Rubin, Donald B},
  journal = {The Annals of Statistics},
  year={2012}
}

@manual{mosek,
   author = "MOSEK ApS",
   title = "MOSEK Optimizer API for Python 10.1.31",
   year = 2024,
   url = "https://docs.mosek.com/latest/pythonapi/index.html"
 }

@article{muller1959note,
  title={A note on a method for generating points uniformly on n-dimensional spheres},
  author={Muller, Mervin E},
  journal={Communications of the ACM},
  volume={2},
  number={4},
  pages={19--20},
  year={1959},
  publisher={ACM New York, NY, USA}
}

@inproceedings{ostrow2016assessment,
  title={The assessment of learning infrastructure (ALI) the theory, practice, and scalability of automated assessment},
  author={Ostrow, Korinn S and Selent, Doug and Wang, Yan and Van Inwegen, Eric G and Heffernan, Neil T and Williams, Joseph Jay},
  booktitle={Proceedings of the sixth international conference on learning analytics \& knowledge},
  pages={279--288},
  year={2016}
}

@article{plackett1946design,
  title={The design of optimum multifactorial experiments},
  author={Plackett, Robin L and Burman, J Peter},
  journal={Biometrika},
  volume={33},
  number={4},
  pages={305--325},
  year={1946},
  publisher={JSTOR}
}

@article{prussing1986principal,
  title={The principal minor test for semidefinite matrices},
  author={Prussing, John E},
  journal={Journal of Guidance, Control, and Dynamics},
  volume={9},
  number={1},
  pages={121--122},
  year={1986}
}

@book{pukelsheim2006optimal,
  title={Optimal design of experiments},
  author={Pukelsheim, Friedrich},
  year={2006},
  publisher={SIAM}
}

@inproceedings{selent2016assistments,
  title={Assistments dataset from multiple randomized controlled experiments},
  author={Selent, Douglas and Patikorn, Thanaporn and Heffernan, Neil},
  booktitle={Proceedings of the Third (2016) ACM Conference on Learning@ Scale},
  pages={181--184},
  year={2016}
}

@article{skippon2016experience,
  title={How experience of use influences mass-market drivers’ willingness to consider a battery electric vehicle: A randomised controlled trial},
  author={Skippon, Stephen M and Kinnear, Neale and Lloyd, Louise and Stannard, Jenny},
  journal={Transportation Research Part A: Policy and Practice},
  volume={92},
  pages={26--42},
  year={2016},
  publisher={Elsevier}
}

@book{trefethen2022numerical,
  title={Numerical linear algebra},
  author={Trefethen, Lloyd N and Bau, David},
  year={2022},
  publisher={SIAM}
}

@inproceedings{xu2015infrastructure,
  title={From infrastructure to culture: A/B testing challenges in large scale social networks},
  author={Xu, Ya and Chen, Nanyu and Fernandez, Addrian and Sinno, Omar and Bhasin, Anmol},
  booktitle={Proceedings of the 21th ACM SIGKDD international conference on knowledge discovery and data mining},
  pages={2227--2236},
  year={2015}
}

@article{zhang2022locally,
  title={Locally optimal design for a/b tests in the presence of covariates and network dependence},
  author={Zhang, Qiong and Kang, Lulu},
  journal={Technometrics},
  volume={64},
  number={3},
  pages={358--369},
  year={2022},
  publisher={Taylor \& Francis}
}

@article{zhang2022min,
  title={Min-max optimal design of two-armed trials with side information},
  author={Zhang, Qiong and Khademi, Amin and Song, Yongjia},
  journal={INFORMS Journal on Computing},
  volume={34},
  number={1},
  pages={165--182},
  year={2022},
  publisher={INFORMS}
}

@article{zhang2024collaborative,
  title={Collaborative Analysis for Paired A/B Testing Experiments},
  author={Zhang, Qiong and Kang, Lulu and Deng, Xinwei},
  journal={Statistica Sinica},
  volume={in press},
  year={2026}
}

\newpage

\appendix

\section{Proof of Proposition 1}

\begin{proof}
We will begin by directly calculating $X^\top V^{-1} X$. First, note that
\begin{align*}
    V  =\begin{bmatrix}
        \sigma_1^2 I_N & \bm{0}_{N\times N} & \ldots & \bm{0}_{N \times N}\\
        \bm{0}_{N\times N} &  \sigma_2^2 I_N & \ldots & \bm{0}_{N \times N}\\
        \vdots & \vdots & \ddots & \vdots \\
        \bm{0}_{N \times N} & \bm{0}_{N \times N} & \ldots & \sigma_K^2 I_N
    \end{bmatrix} + \tau^2 \begin{bmatrix}
        I_N\\
        \vdots\\
        I_N
    \end{bmatrix}\begin{bmatrix}
        I_N & \ldots & I_N
    \end{bmatrix}.
\end{align*}
By applying the Woodbury formula \citep{higham2002accuracy}, we can see that
\begin{align*}
    V^{-1} = \begin{bmatrix}
        \sigma_1^{-2} I_N & \bm{0}_{N\times N} & \ldots & \bm{0}_{N \times N}\\
        \bm{0}_{N\times N} &  \sigma_2^{-2} I_N & \ldots & \bm{0}_{N \times N}\\
        \vdots & \vdots & \ddots & \vdots \\
        \bm{0}_{N \times N} & \bm{0}_{N \times N} & \ldots & \sigma_K^{-2} I_N
    \end{bmatrix} - \Big(\tau^{-2} + \sum_{\ell=1}^K\sigma_\ell^{-2}\Big)^{-1} \begin{bmatrix}
        \sigma_1^{-2}I_N\\
        \vdots\\
        \sigma_K^{-2}I_N
    \end{bmatrix}\begin{bmatrix}
        \sigma_1^{-2}I_N & \ldots & \sigma_K^{-2}I_N
    \end{bmatrix}.
\end{align*}
So then direct computation of $X^\top V^{-1} X$ yields
\begin{align*}
    X^\top V^{-1}X = G - \big(\tau^{-2} + \sum_{\ell=1}^K \sigma_\ell^{-2}\big)^{-1} H
\end{align*}
where 
\begin{align*}
    G = \begin{bmatrix}
        \sigma_1^{-2}\bm{x}_1^\top\bm{x}_1 & 0 & \ldots & 0 & \sigma_1^{-2}\bm{x}_1Z & \bm{0}_{1\times p} & \ldots & \bm{0}_{1 \times p}\\
        0 & \sigma_2^{-2}\bm{x}_2^\top\bm{x}_2 & \ldots & 0 & \bm{0}_{1 \times p}  & \sigma_2^{-2}\bm{x}_2Z & \ldots & \bm{0}_{1 \times p}\\
        \vdots & \vdots & \ddots & \vdots & \vdots & \vdots & \ddots &\vdots \\
        0 & 0 & \ldots & \sigma_K^{-2}\bm{x}_K^\top \bm{x}_K & \bm{0}_{1 \times p} & \bm{0}_{1 \times p} & \ldots & \sigma_K^{-2}\bm{x}_K^\top Z\\
        \sigma_1^{-2}Z^\top \bm{x}_1 & \bm{0}_{p \times 1} & \ldots & \bm{0}_{p \times 1} & \sigma_1^{-2}Z^\top Z & \bm{0}_p & \ldots & \bm{0}_p\\
        \bm{0}_{p \times 1} & \sigma_2^{-2}Z^\top \bm{x}_2 & \ldots & \bm{0}_{p \times 1} & \bm{0}_p & \sigma_2^{-2}Z^\top Z & \ldots & \bm{0}_p\\
        \vdots & \vdots & \ddots & \vdots & \vdots & \vdots & \ddots &\vdots \\
        \bm{0}_{p \times 1} & \bm{0}_{p \times 1} & \ldots & \sigma_K^{-2}Z^\top \bm{x}_K & \bm{0}_p & \bm{0}_p & \ldots & \sigma_K^{-2}Z^\top Z
    \end{bmatrix}
\end{align*}
and
\begin{align*}
    H = \begin{bmatrix}
        \sigma_1^{-2}\sigma_1^{-2}\bm{x}_1^\top \bm{x}_1 & \sigma_1^{-2}\sigma_2^{-2}\bm{x}_1^\top \bm{x}_2 &\ldots & \sigma_1^{-2}\sigma_K^{-2} \bm{x}_1^\top \bm{x}_K & \sigma_1^{-2}\sigma_1^{-2}\bm{x}_1^\top Z & \sigma_1^{-2}\sigma_2^{-2}\bm{x}_2^\top Z & \ldots & \sigma_1^{-2}\sigma_K^{-2}\bm{x}_1^\top Z\\
        \sigma_2^{-2}\sigma_1^{-2}\bm{x}_2^\top \bm{x}_1 & \sigma_2^{-2}\sigma_2^{-2}\bm{x}_2^\top \bm{x}_2 &\ldots & \sigma_2^{-2}\sigma_K^{-2} \bm{x}_2^\top \bm{x}_K & \sigma_2^{-2}\sigma_1^{-2}\bm{x}_2^\top Z & \sigma_2^{-2}\sigma_2^{-2}\bm{x}_2^\top Z & \ldots & \sigma_2^{-2}\sigma_K^{-2}\bm{x}_2^\top Z\\
        \vdots & \vdots & \ddots & \vdots & \vdots & \vdots & \ddots &\vdots \\
        \sigma_K^{-2}\sigma_1^{-2}\bm{x}_K^\top \bm{x}_1 & \sigma_K^{-2}\sigma_2^{-2}\bm{x}_K^\top \bm{x}_2 &\ldots & \sigma_K^{-2}\sigma_K^{-2} \bm{x}_K^\top \bm{x}_K & \sigma_K^{-2}\sigma_1^{-2}\bm{x}_K^\top Z & \sigma_K^{-2}\sigma_2^{-2}\bm{x}_K^\top Z & \ldots & \sigma_K^{-2}\sigma_K^{-2}\bm{x}_K^\top Z\\
        \sigma_1^{-2}\sigma_1^{-2}Z^\top \bm{x}_1 &  \sigma_1^{-2}\sigma_2^{-2}Z^\top \bm{x}_2 &\ldots & \sigma_1^{-2}\sigma_K^{-2} Z^\top \bm{x}_K & \sigma_1^{-2}\sigma_1^{-2}Z^\top Z & \sigma_1^{-2}\sigma_2^{-2}Z^\top Z & \ldots & \sigma_1^{-2}\sigma_K^{-2}Z^\top Z\\
        \sigma_2^{-2}\sigma_1^{-2}Z^\top \bm{x}_1 & \sigma_2^{-2}\sigma_2^{-2}Z^\top \bm{x}_2 &\ldots & \sigma_2^{-2}\sigma_K^{-2} Z^\top \bm{x}_K & \sigma_2^{-2}\sigma_1^{-2}Z^\top Z & \sigma_2^{-2}\sigma_2^{-2}Z^\top Z & \ldots & \sigma_2^{-2}\sigma_K^{-2}Z^\top Z\\
        \vdots & \vdots & \ddots & \vdots & \vdots & \vdots & \ddots &\vdots \\
        \sigma_K^{-2}\sigma_1^{-2}Z^\top \bm{x}_1 & \sigma_K^{-2}\sigma_2^{-2}Z^\top \bm{x}_2 &\ldots & \sigma_K^{-2}\sigma_K^{-2} Z^\top \bm{x}_K & \sigma_K^{-2}\sigma_1^{-2}Z^\top Z & \sigma_K^{-2}\sigma_2^{-2}Z^\top Z & \ldots & \sigma_K^{-2}\sigma_K^{-2}Z^\top Z\\
    \end{bmatrix}.
\end{align*}
Letting $c = 1 + \tau^2 \sum_{\ell=1}^K\sigma_\ell^{-2}$, $Q_j = \sigma_j^{-2}(c-\tau^2\sigma_j)$, and $R_{j,j'} = -\tau^2 \sigma_j\sigma_{j'}$ we then have that 
\begin{align*}
       X^\top V^{-1}X =  \frac{1}{c}\begin{bmatrix}
        Q_1\bm{x}_1^\top \bm{x}_1 & R_{1,2}\bm{x}_1^\top \bm{x}_2 &\ldots & R_{1,K} \bm{x}_1^\top \bm{x}_K & Q_1\bm{x}_1^\top Z & R_{1,2}\bm{x}_2^\top Z & \ldots & R_{1,K}\bm{x}_1^\top Z\\
        R_{2,1}\bm{x}_2^\top \bm{x}_1 & Q_2\bm{x}_2^\top \bm{x}_2 &\ldots & R_{2,K} \bm{x}_2^\top \bm{x}_K & R_{2,1}\bm{x}_2^\top Z & Q_2\bm{x}_2^\top Z & \ldots & R_{2,K}\bm{x}_2^\top Z\\
        \vdots & \vdots & \ddots & \vdots & \vdots & \vdots & \ddots &\vdots \\
        R_{K,1}\bm{x}_K^\top \bm{x}_1 & R_{K,2}\bm{x}_K^\top \bm{x}_2 &\ldots & Q_K \bm{x}_K^\top \bm{x}_K & R_{K,1}\bm{x}_K^\top Z & R_{K,2}\bm{x}_K^\top Z & \ldots & Q_K\bm{x}_K^\top Z\\
        Q_1 Z^\top \bm{x}_1 &  R_{1,2}Z^\top \bm{x}_2 &\ldots & R_{1,K} Z^\top \bm{x}_K & Q_1 Z^\top Z & R_{1,2}Z^\top Z & \ldots & R_{1,K}Z^\top Z\\
        R_{2,1}Z^\top \bm{x}_1 & Q_2Z^\top \bm{x}_2 &\ldots & R_{2,K} Z^\top \bm{x}_K & R_{2,1}Z^\top Z & Q_2Z^\top Z & \ldots & R_{2,K}Z^\top Z\\
        \vdots & \vdots & \ddots & \vdots & \vdots & \vdots & \ddots &\vdots \\
        R_{K,1}Z^\top \bm{x}_1 & R_{K,2}Z^\top \bm{x}_2 &\ldots & Q_K Z^\top \bm{x}_K & R_{K,1}Z^\top Z & R_{K,2}Z^\top Z & \ldots & Q_KZ^\top Z\\
    \end{bmatrix}.
\end{align*}
Note that $X^\top V^{-1} X$ can be partitioned into a $2\times 2$ block matrix as 
\begin{align*}
    X^\top V^{-1} X = \frac{1}{c}\begin{bmatrix}
        M_1 & M_2\\
        M_2^\top & M_3
    \end{bmatrix},
\end{align*}
where $M_1$ corresponds to the $K\times K$ upper left submatrix of $X^\top V^{-1} X$ involving terms $\bm{x}_j^\top \bm{x}_{j'}$, $M_2$ is the upper right submatrix of $X^\top V^{-1} X$ involving terms $\bm{x_j}^\top Z$, and $M_3$ is the lower right submatrix of $X^\top V^{-1} X$ involving terms $Z^\top Z$. 

Recalling that $X^\top V^{-1} X$ is the precision matrix of $[\hat{\bm{\beta}}^\top, \hat{\bm{\gamma}}^\top]^\top$, it follows that the covariance matrix of $\hat{\bm{\beta}}$ is given by the upper left $K \times K$ submatrix of $(X^\top V^{-1} X)^{-1}$. It follows from block-matrix inversion that
\begin{align*}
    \text{Cov}(\hat{\bm{\beta}}) = c (M_1-M_2 M_3^{-1}M_2^\top)^{-1}.
\end{align*}
Thus the precision of $\hat{\bm{\beta}}$ is simply $\frac{1}{c}(M_1-M_2 M_3^{-1}M_2^\top)$.
Letting
\begin{align*}
    S &= \begin{bmatrix}
        Q_1 & R_{1,2} & \ldots & R_{1,K}\\
        R_{2,1} & Q_2 & \ldots & R_{2,K}\\
        \vdots & \vdots & \ddots & \vdots\\
        R_{K,1} & R_{K,2} & \ldots & Q_K
    \end{bmatrix}\\
    &=c\begin{bmatrix}
        \sigma_1^{-2} & 0 & \ldots & 0\\
        0 & \sigma_2^{-2} & \ldots & 0\\
        \vdots & \vdots & \ddots & \vdots\\
        0 & 0 & \ldots & \sigma_K^{-2}
    \end{bmatrix} - \tau^2 \begin{bmatrix}
        \sigma_1^{-2} \\
        \vdots \\
        \sigma_K^{-2}
    \end{bmatrix}\begin{bmatrix}
        \sigma_1^{-2} & \ldots & \sigma_K^{-2}
    \end{bmatrix},
\end{align*}
we can easily see that $M_3 = S \otimes (Z^\top Z)$, where $\otimes$ denotes the Kronecker product. Under our assumption that $Z$ is full rank, and assuming that $S$ is invertible (which it is when $\sigma_j \not= 0$ for all $j$), it follows that $M_3^{-1} = (S \otimes (Z^\top Z))^{-1} = S^{-1} \otimes (Z^\top Z)^{-1}$ from the inversion property of the Kronecker product. Through another use of the Woodbury formula, we can show that 
\begin{align*}
    S^{-1} = \frac{1}{c}\begin{bmatrix}
        (\sigma_1^{2} + \tau^2) & \tau^2 & \ldots & \tau^2\\
        \tau^2 & (\sigma_2^2 + \tau^2) & \ldots & \tau^2\\
        \vdots & \vdots & \ddots & \vdots\\
        \tau^2 & \tau^2 & \ldots & (\sigma_K^2 + \tau^2)
    \end{bmatrix}.
\end{align*}
Using this information, along with the observation that $(\sigma_j^2 + \tau^2)Q_j + \tau^2 \sum_{\ell \not= j} R_{\ell,j} = c$ and $\sigma_j^2 R_{j,j'} + \tau^2 Q_{j'} + \tau^2 \sum_{\ell \not= j'} R_{\ell,j'} = 0$, direct computation of $\frac{1}{c}(M_1 - M_2 M_3^{-1}M_2^\top)$ yields the precision matrix of $\hat{\bm{\beta}}$.
\end{proof}

\section{Proof of Theorem \ref{thm:upper}}

According to the Hadamard's inequality \citep{hadamard1893resolution}, we have that
\[
 \text{det}\Big(P(\hat{\bm{\beta}})\Big)\leq \frac{1}{c^K}\prod^K_{j=1} Q_j\mathbf{x}_j^\top P_{Z^\perp} \mathbf{x}_j,
\]
where the equality is attained if the off-diagonal entries all equal to zero, i.e., 
\[
\mathbf{x}_j^\top P_{Z^\perp} \mathbf{x}_{j'}=0
\]
for any two distinct $j, j'\in\{1, \ldots, K\}$. 

Further, we have that $\mathbf{x}_j^\top P_{Z^\perp} \mathbf{x}_j\leq N$, and 
equality is attained if $\mathbf{x}_j$ is perfectly covariate-balanced, i.e.,  $Z^\top\mathbf{x}_j=\mathbf{0}$. Under perfect covariate-balancing for all $k\in \{1, \ldots, K\}$, we have that $\mathbf{x}_j^\top P_{Z^\perp} \mathbf{x}_j'=0$ is simplified to mutual orthogonality between $\mathbf{x}_1, \ldots, \mathbf{x}_K$. This completes the proof.

\section{Proof of Theorem \ref{thm:var}}

For notational convenience, we use the case of $k=1$ to prove the result.

For the lower bound, 
\[
\mathrm{Var}(\hat\beta_1)=c\left[Q_1 \mathbf{x}_1^\top P_{Z^\perp} \mathbf{x}_1-\bm{x}^\top_1 P_{Z^\perp} R_{1,-1}R^{-1}_{-1, -1}R^\top_{1,-1} P_{Z^\perp} \bm{x}_1\right]^{-1}\geq c\left[Q_k \mathbf{x}_1^\top P_{Z^\perp} \mathbf{x}_1\right]^{-1}\geq \frac{c}{Q_1 N},
\]
where the equality holds if $Z^\top\bm x_1=0$ and
$\bm{x}^\top_1 P_{Z^\perp} R_{1,-1}=\bm{x}^\top_1 P_{Z^\perp}[R_{1,2}\bm x_2, \ldots, R_{1,K}\bm x_K]=0$.

For the upper bound, $\hat\beta_1$ is the best linear unbiased estimator of $\beta_1$. Therefore, the variance of $\hat\beta_1$
is no larger than any other linear unbiased estimator of $\beta_1$. Therefore, the conclusion holds. 
Suppose that all the experiments have the same design, then 
\[
P(\hat{\bm{\beta}})=\mathbf{x}_1^\top P_{Z^\perp} \mathbf{x}_1 \left(\mathrm{diag}(\bm s)-\frac{\tau^2}{c}\bm s\bm s^\top\right),
\]
where $\bm s=[\sigma^{-2}_1, \ldots, \sigma^{-2}_K]^\top$. Therefore, $\mathrm{Var}(\hat\beta_1)$ is given by the first diagonal entry of the matrix $[P(\hat{\bm{\beta}})]^{-1}$. Note that
\[
\left(\mathrm{diag}(\bm s)-\frac{\tau^2}{c}\bm s\bm s^\top\right)^{-1}
=\mathrm{diag}(\bm s)^{-1}+\tau^2\bm 1\bm 1^\top.
\]
Therefore 
\[
\mathrm{Var}(\hat\beta_1)=\frac{\sigma^2_1+\tau^2}{\mathbf{x}_1^\top P_{Z^\perp} \mathbf{x}_1}=\mathrm{Var}(\tilde\beta_1).
\]

\section{Additional Case-Study Results for $\sigma=0.1$} \label{app: sigma_0_1_results}
\begin{figure}[p]
    \centering
    \captionsetup{font=small}
    \includegraphics[width=\textwidth]{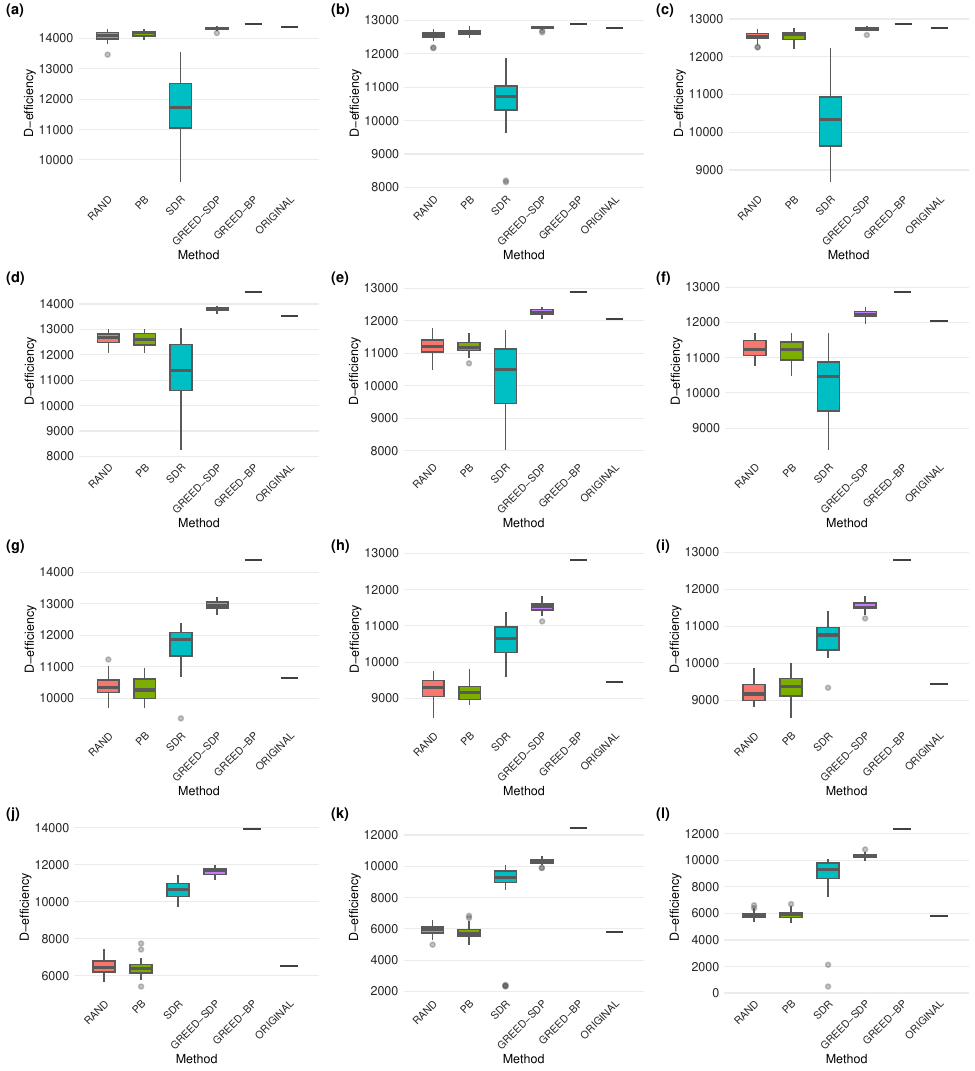}
    \caption{D-efficiencies for $\sigma=0.1$. Rows, from top to bottom, use total covariate dimensions $p=5,25,55,105$, including the intercept; columns, from left to right, use $\tau=0.1,1,10$. Panels (a)--(l) are ordered across rows. Boxplots summarize 30 designs per randomized method; horizontal segments show the single GREED-BP and ORIGINAL values.}
    \label{fig:app_deff_sigma01_combined}
\end{figure}

Here we present the D-efficiency results for the case where $\sigma=0.1$.
Figure \ref{fig:app_deff_sigma01_combined} gives the D-efficiency boxplots for the original ASSISTments design and the five constructed designs under $\tau\in\{0.1,1,10\}$ for four covariate settings.
As in Section \ref{subsec: casestudyresults}, each boxplot summarizes 30 generated designs for a randomized method, while GREED-BP and ORIGINAL are represented by single horizontal segments.

The main conclusions are consistent with the $\sigma=1$ results in Section \ref{subsec: casestudyresults}. 
When $\sigma=0.1$, the D-efficiency values are generally higher across all methods and settings, which is expected since the response noise is smaller.
The relative performance of the methods is similar to the $\sigma=1$ case:
GREED-BP has the highest observed D-efficiency in every displayed setting, and GREED-SDP has the second-highest mean D-efficiency for $p=25$, $55$, and $105$. 
SDR has intermediate mean D-efficiency for $p=55$ and $p=105$, but the lowest mean D-efficiency for $p=5$ and $p=25$. 
RAND and PB lose more efficiency as $p$ increases.


\end{document}